\documentclass[conference]{IEEEtran}

\ifCLASSOPTIONcompsoc
  \usepackage[nocompress]{cite}
\else
  \usepackage{cite}
\fi

\ifCLASSINFOpdf
\usepackage[pdftex]{graphicx}
\graphicspath{{./}}
\DeclareGraphicsExtensions{.pdf,.jpeg,.png}
\else
\fi

\usepackage{url}

\usepackage{listings}
\usepackage{xcolor}
\usepackage{hyphenat}
\usepackage{bytefield}
\usepackage{subcaption}
\usepackage{graphicx}
\usepackage{comment}
\usepackage{booktabs}
\usepackage{pifont}
\usepackage{balance}

\newcommand{\ovs}{\textit{OvS}}

\newcommand{\of}{OpenFlow}
\newcommand{\os}{OpenStack}

\newcommand{\shortmpls}{Short Shim Attack}
\newcommand{\longmpls}{Long Shim Attack}

\newcommand{\grsec}{grsecurity}

\newcommand{\kash}[1]{\textit{\textcolor{red}{[kash]: #1}}} 

\makeatletter
\def\lst@makecaption{%
  \def\@captype{table}%
  \@makecaption
}
\makeatother

\begin{document}

\title{Security and Dependability Issues in Open vSwitch}
\title{Dataplane Vulnerabilities in the Cloud}
\title{Give me Your Cloud and I will Own it}
\title{Exploiting Network Function Vulnerabilities to Bring Down the Cloud}
\title{Network Function Vulnerabilities (NFV): Compromising the Cloud}
\title{{\Large Network Function Vulnerability (NFV):}\\ Compromising the Cloud From a (Data) Plane}

\title{Compromising the Cloud From a (Data) Plane}

\title{Compromising the Cloud from a\\ 
(Data) Plane is Cheap}

\title{Soft-Switches: From data plane security assumptions to compromised cloud systems}

\title{Compromising Virtualized Cloud Systems\\via the Data Plane}
\title{Reins to the Cloud:\\Compromising Cloud Systems via the Data Plane}

\author{
\IEEEauthorblockN{Kashyap Thimmaraju\IEEEauthorrefmark{1}\IEEEauthorrefmark{2}, Bhargava Shastry\IEEEauthorrefmark{1}\IEEEauthorrefmark{2}\\ Tobias Fiebig\IEEEauthorrefmark{1},
Felicitas Hetzelt\IEEEauthorrefmark{1}\IEEEauthorrefmark{2}\\ Jean-Pierre Seifert\IEEEauthorrefmark{1}\IEEEauthorrefmark{2}, Anja Feldmann\IEEEauthorrefmark{1} and Stefan Schmid\IEEEauthorrefmark{1}\IEEEauthorrefmark{3}}
\\

\IEEEauthorblockA{\IEEEauthorrefmark{1}TU Berlin \\
kashyap.thimmaraju@sec.t-labs.tu-berlin.de, bshastry@sec.t-labs.tu-berlin.de, tobias@inet.tu-berlin.de\\ file@sec.t-labs.tu-berlin.de, anja@inet.tu-berlin.de}
\\
\IEEEauthorblockA{\IEEEauthorrefmark{2}Telekom Innovation Laboratories\\
jean-pierre.seifert@telekom.de}
\\
\IEEEauthorblockA{\IEEEauthorrefmark{3}Aalborg University\\
schmiste@cs.aau.dk
}

}


\maketitle

\begin{abstract} 
Virtual switches have become
popular among cloud operating systems to interconnect virtual machines
in a more flexible manner. However,  
this paper demonstrates that virtual switches
introduce new attack surfaces  in cloud setups, whose 
effects can be disastrous. Our analysis shows that these vulnerabilities are caused by: (1)  inappropriate security assumptions (privileged virtual switch execution in kernel and user space),
(2) the logical centralization of such networks (e.g., OpenStack or SDN),  
(3) the presence of bi-directional communication channels between data plane systems and the
centralized controller, and
(4) non-standard protocol parsers.

Our work highlights the need to accommodate the data plane(s) in
our threat models. In particular, it forces us to revisit today's assumption that the data
plane can only be compromised by a sophisticated attacker: we show that
compromising the data plane of modern computer networks can actually be
performed by a very simple attacker with limited resources only and at low cost
(i.e., at the cost of renting a virtual machine in the Cloud).
As a case study, we fuzzed only 2\% 
of the code-base of a production quality virtual switch's packet
processor (namely OvS), identifying 
serious vulnerabilities leading to unauthenticated remote
code execution. In particular, we present the ``rein worm'' which allows us to fully compromise test-setups in less than 100 seconds. 
We also evaluate the performance overhead of existing mitigations such as ASLR,
PIEs, and unconditional stack canaries on OvS.
We find that while applying these
countermeasures in kernel-space incurs a significant overhead,
in user-space the performance overhead is negligible.
\end{abstract}

\IEEEpeerreviewmaketitle

\section{Introduction}
\label{sec:intro}
Computer networks are becoming increasingly programmable and virtualized:
software switches and virtualized network functions run on commodity hardware.
The virtualization of such packet processing functions facilitates a
flexible and faster definition and deployment of new network functions,
essentially using a simple software update.  This is also attractive from a
costs perspective~\cite{ossurvey}: Today's computer networks host a large
number of expensive, complex, and inflexible hardware routers, middleboxes and
appliances (e.g., firewalls, proxies, NATs, WAN optimizers, etc.). The latter
can be in the order of the number of
routers~\cite{someone,manifesto,opennf,opensdwn,martins2014clickos}.  Moreover,
technological advances as well as the quickly increasing core density per host
processor render it possible to perform even resource intensive data plane
functions \emph{at line rate} on commodity servers, i.e., at hundreds of
Gbps~\cite{blake2009survey}.  The performance of software switching can be
further improved using hardware-offloading which is gaining traction.
Accordingly, so-called \emph{virtual switches} are becoming popular,
especially in datacenters~\cite{179731,reading}.

Hand-in-hand with the increasing virtualization and programmability of
networked systems comes an increasing \emph{centralization}: the control over
network elements is outsourced and consolidated to a logically centralized
control plane (e.g., the controller of cloud operating systems such as
OpenStack~\cite{osnetworkguide}).
 Hence the
logically centralized perspective can significantly simplify reasoning about
orchestrating and operating distributed systems.

The canonical example which combines these two trends is Software-Defined Networks
(SDNs): in SDN, the control over the data plane elements (typically the
OpenFlow switches) is outsourced to a logically centralized software (the
so-called controller) running on a server platform. The controller interacts
with the OpenFlow switches via the OpenFlow API using a bidirectional
communication channel.  Especially in datacenters, virtual switches (such as Open vSwitch,
Cisco Nexus 1000V, VMware's vSwitch) are popular for the flexibilities
they provide in terms of network virtualization~\cite{179731} (e.g., to
control, police, and dynamically handle virtual machine traffic), as well for
their simple edge-based deployment model~\cite{greenberg2015sdn,commodity}.

\begin{figure}[t!]
	\centering
    \includegraphics[trim=0.0cm 0.0cm 0.0cm 0.0cm,clip=true,width=.9\columnwidth]{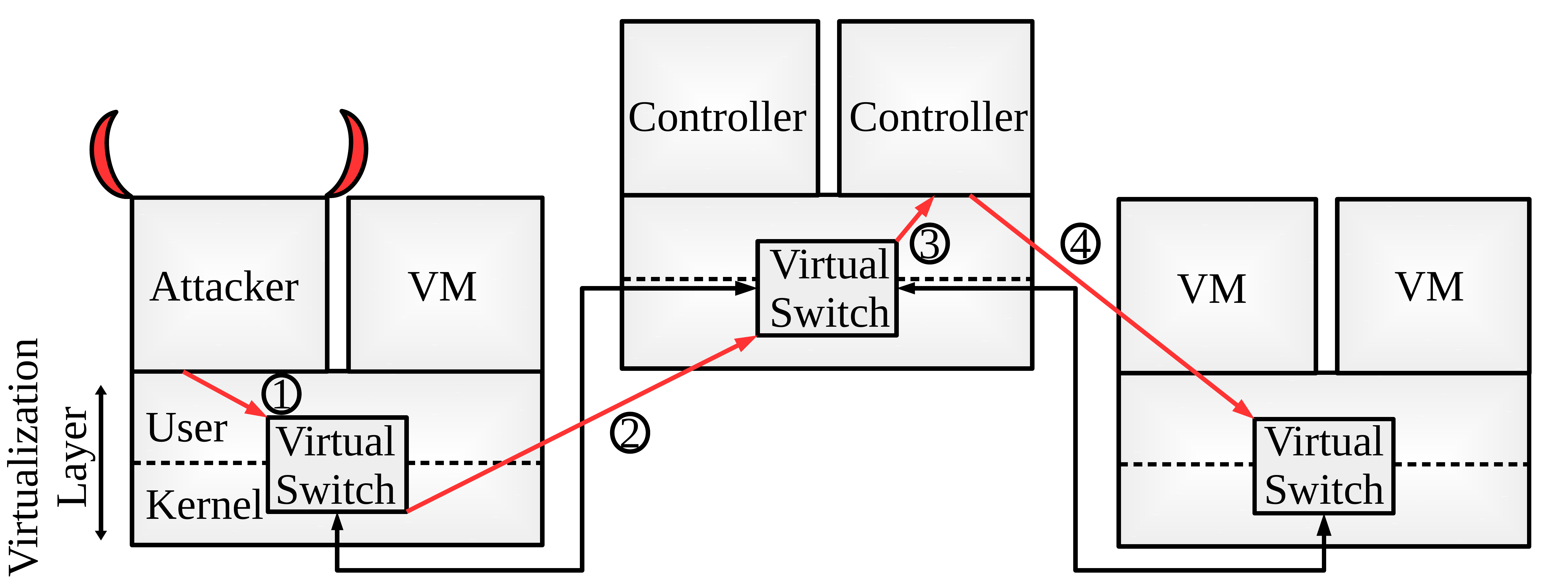}
    \caption{Overview: This paper explores the effect of
    vulnerabilities introduced by virtual switches in the cloud. 
    We consider a datacenter consisting of multiple 
    virtualized (commodity) servers hosting different virtual machines, 
    interconnected by virtual switches. The control over network elements is 
    outsourced to a logically centralized controller 
    (e.g., OpenStack or SDN), interacting with the data plane 
    via bidirectional communication channels. The virtual switch is 
    situated in both user- and kernel-space. We show that an attacker VM (\emph{on the left}),
can exploit a vulnerable switch to compromise the server. From there, 
the attacker can compromise the controller (server) and then
manipulate the virtual switch (\emph{on the right}).}
    \label{fig:cloudComputing}
\end{figure}

\subsection{Our Contributions}

This paper shows that the virtualization and centralization
of modern computer networks 
introduce
new attack surfaces and
exploitation opportunities with the data plane. 
In particular, we present a security analysis
of virtual switches. We show that even a simple, low-resource attacker
can exploit data plane
elements, and thereby
compromise critical 
cloud software services,
obtain direct access to the 
SDN southbound
interface, 
or violate network security policies (cf.~Figure~\ref{fig:cloudComputing}).

This differs from prior research on 
SDN security, which has primarily focussed on the
control plane~\cite{porras2015securing,fresco}.
Furthermore, attacks from the data plane have often been
assumed to require significant resources~\cite{schuchard2010losing}
or state-level compromise (by controlling the
vendor and/or its supply-chain)~\cite{snowdencisco}.
In contrast, 
we in this paper show that attacks on and from
the data plane of modern computer networks can actually be
performed by a \emph{simple attacker}. Thus forcing us to revise
today's threat models.

Hence, our key conceptual contributions are:
\begin{enumerate}
\item 
We point out and analyze novel vulnerabilities and attack opportunities
arising in the context of virtual switches running on commodity server 
systems. In particular, we show that, via the data plane and by 
inappropriate privileges
(running as root in user-space),
compared to (non-virtualized, non-centralized) traditional networks,  
an attacker can cause significant harm and
compromise essential 
cloud services. 

\item 
We show that 
it is
cheap to launch such attacks in the context
of virtualized networks:
an unsophisticated attacker simply needs  
access to a Virtual Machine (VM) in the datacenter to perform
our exploit.
There is no need, e.g., to install (hardware) backdoors to compromise
a switch.
\end{enumerate}

To highlight the severity of the problem, 
we fuzzed the packet processor of the 
state-of-the-art 
production quality virtual
switch, namely Open vSwitch (OvS).
OvS is the default virtual switch in OpenStack, Xen, Pica8, 
among others and is shipped as a kernel module
in many Linux distributions such
as Ubuntu, RedHat, OpenSuse, etc.
Fuzzing a small fraction
of the code-base (less than 2\%)
was sufficient to uncover 
exploitable software vulnerabilities in the packet
processing engine of OvS.
The vulnerabilities can be exploited for remote code execution.
In particular, we demonstrate how the 
\emph{Rein worm} which, starting from a VM within an OpenStack cloud, 
can first compromise the
entire host operating system of the underlying server. 
From there, the rein worm
propagates 
to the controller and subsequently compromises the controller's
operating system. 
Once at the
controller, the rein worm spreads to all the other servers
that are connected to the controller (see~Figure~\ref{fig:cloudComputing}). 
At each stage, 
the rein worm compromises confidentialy, integrity, and availability of the
respective servers.
We experimentally demonstrate that the rein worm can compromise an
OpenStack deployment of 100 servers
in less than 100 seconds.

We complement our vulnerability analysis
by studying possible countermeasures. 
Our empirical performance study shows that 
software-based
countermeasures such as stack canaries and
position independent executables 
do not affect 
the forwarding
performance (throughput and latency)
of the slow path of~OvS by much.
However, the use of grsecurity kernel patches~\cite{PaX01}
does entail a non-trivial 
performance overhead.
We suggest using such countermeasures,
especially for userland applications 
in production environments.
Moreover, we believe that our measurement study
constitutes an independent contribution of this paper:
we are unaware of studies
targeted at measuring the performance overhead of different 
software-based security protection mechanisms for virtual switches such as~OvS. 

\subsection{Ethical Considerations and Impact}

To avoid disrupting the normal operation
of businesses, we verified our findings on our
own infrastructure. However, we 
have disclosed our findings in a secure manner to the OvS
team who have propagated
the fixes downstream. Ubuntu, Redhat, Debian, Suse, Mirantis, and other
stakeholders have applied the patch(es).
Some of the bugs have also been published under CVE-2016-2074.

Indeed, we believe that the specific remote code execution
vulnerability identified in this paper is of practical relevance.
Virtual switches such as OvS
are quite popular among cloud operating systems (virtual management
systems) such as OpenStack, oVirt, OpenNebula, etc.
According to the OpenStack Survey 2016~\cite{ossurvey},
over 60\% OvS deployments are in production use
and over one third of 1000+ core clouds use 
OvS
(directional data only). 
The identified vulnerability is also relevant 
because it can be leveraged to harm essential services
of the cloud operating system,
including, e.g.: 
managed compute resources (hypervisors and guest VMs),
image management (the images VMs use for boot-up),
block storage (data storage),
network management (virtual networks between hypervisors and guest VMs),
for the dashboard and web UI (in order to manage the various resources from a centralized location), identity managment (of the adminstrators and tenants),
etc.

While our case study focuses on SDN, 
the relevance of our threat model is more general.
The importance of our threat model is also likely to increase 
with the advent of 5G networks~\cite{6812298} and increasing deployment of
Network Function Virtualization 
(NFV)~\cite{martins2014clickos}
or protocol independent packet processing systems like P4~\cite{bosshart2014p4,barefoot}.

\subsection{Organization}

The remainder of this paper is organized as follows.
We provide background information required
to comprehend the rest of this paper in
Section~\ref{sec:background}.
Section~\ref{sec:compromise} 
introduces, discusses, and analyses the vulnerabilities
identified in this paper, and derives our threat model accordingly.
Section~\ref{sec:proofofconcept} presents a proof-of-concept
and case study of 
our threat model and attacks with
OvS in OpenStack.
Subsequently, in Section~\ref{sec:eval}, we 
describe our empirical findings on
the forwarding performance of OvS
with software countermeasures.
In Section~\ref{sec:discussion}
we discuss security concepts
and packet processing in a broad context.
After reviewing related work in
Section~\ref{sec:literature},
we conclude our contribution in
Section~\ref{sec:conclusion}.

\section{Background}
\label{sec:background}
This section provides the necessary background 
and terminology required to understand
the remainder of this paper.

\subsection{Centralized Cloud and Network Control}

Modern cloud operating systems such as OpenStack, OpenNebula, etc.~are
designed for (logically) centralized network control and global
visibility. Data plane isolation is typically ensured using 
separate physical/logical networks (guest, management
and external) and tunneling technologies
such as VLAN, GRE, VXLAN, MPLS, etc.
A cloud network generally comprises of
a physical network consisting of
physical switches interconnecting
virtualized servers and an overlay (virtual)
network
interconnecting
the VMs
and their servers.
The centralized control 
is attractive as it reduces the
operational cost and complexity
of managing the cloud network.
It also provides flexibilities for
managing and using cloud services,
including VM migration.

Centralized network control in the cloud
can be offered in different ways,
using the controller of the cloud solution itself
or
using a dedicated SDN controller.
In the former scenario, the controller
can use its own data plane
to communicate with the data plane of the
servers. In the latter scenario,
the SDN controller directly
communicates with the data plane
of the server(s).
Additionally, the SDN controller can also be used
to manage the physical switches of
the cloud network.

OpenFlow is the de facto standard SDN
protocol today. Via the OpenFlow API, the controller 
can add, remove, update and monitor flow tables and 
their flows.

\subsection{Virtual Switches}

The network data plane(s) 
can either be distributed across the virtualized
servers or across physical (hardware) switches.
OvS, VMware vSwitch and Cisco Nexus 1000V are
examples of the former and are commonly
called \emph{virtual switches}, while
Cisco VN-Link~\cite{ciscoVNlink} and 
Virtual Ethernet Port Aggregator (VEPA)~\cite{kamath2010edge}
are examples of the
latter.

Virtual switches have the advantage that 
inter-VM traffic within a server
does not have to leave the server.
The main purpose of the physical switches is
to offer line rate communication.
The downside, however is that 
the hypervisor or host OS
increases its attack surface, 
thereby
reducing the security of the server.
The performance overhead of
software-only switching (e.g., OvS)
can be alleviated
by hardware-offloading features: 
While such features were
previously only available in expensive proprietary
networking equipment, they are currently gaining
traction. 
Pettit et al.~\cite{pettit2010virtual}
showed that
the performance of OvS{} and VEPA are
comparable when
executing on a remote bare-metal server. 
OvS performs better in case of large
transfers at high rates when
executing on the same server.

The requirements and operating environment of virtual switches
differ signifcantly from those of traditional
network appliances in terms of \emph{resource sharing}
and
\emph{deployment}.
In contrast to traditional network appliances,
virtual switches need to be general enough to
perform well on different platforms,
without the luxury
of specialization~\cite{pfaff2015design}.
Moreover,
virtual switches are typically deployed \emph{at the edge} 
of the network, 
sharing fate, resources, and workloads with the hypervisor and VMs.

The virtual switch broadly comprises of two main components:
management/configuration and forwarding.
These components may be spread across the system. That is,
they may exist as separate processes and/or reside
in user-space and/or kernel-space.
The mangement and configuration 
component deals with administering the virtual
switch (e.g., configuring ports, policies, etc.).
The forwarding component 
is usually based on a sequential (and circular) packet processing pipeline.
The pipeline begins with processing a packet's header
information to extract relevant information
that is used to perform a (flow) table lookup which is generally the
second stage in the pipeline.
The result of the lookup determines the fate of the packet which is
the last stage in the pipeline. Note that the final stage
may result in sending the packet back to the first stage.
We argue that the first stage in the pipeline is
the most vulnerable to an attack for the following reasons:
it accepts arbitrary packet formats,
it is directly influenced by the attacker, and
it typically exists in kernel- and user-space.

\subsection{Open vSwitch}
\label{sec:virtualSwitches}

Open vSwitch (OvS)~\cite{Casado:2010:VNF:1921151.1921162,pfaff2015design,pfaff2009extending,179731} 
is a popular open source and multi-platform virtual switch,
meeting the high performance
requirements of production environments 
as well as the 
programmability demanded by
network virtualization. 
OvS is the default virtual switch for OpenStack, Xen, 
Pica8 and an array of other software,
and primarily seen as an 
SDN switch.
OvS's database can be managed
by the controller via a TCP connection
using the ovsdb protocol.

At the heart of the OvS design are
two
forwarding paths: the
slow path which is a
userspace daemon (\emph{ovs-vswitchd})
and the
fast path which is a
datapath kernel module (\emph{openvswitch.ko}).
OvS also has the capability
to use a hardware switch for the fast path (e.g., Pica8).
Only \emph{ovs-vswitchd} 
can install rules and associated actions on how to handle packets
in the fast path, 
e.g., forward packets to ports or tunnels, 
modify 
packet headers, sample packets, drop packets, etc. 
When a packet does not match a rule in the fast path, the packet is delivered
to \emph{ovs-vswitchd}, which can then determine, in 
userspace, how to handle the packet, 
and then pass it back to the datapath kernel module specifying 
the desired handling.

To improve performance for similar future packets, 
flow caching 
is used. OvS supports two
main cache flavors: \emph{microflow cache} and 
\emph{megaflow cache}. Oversimplifying things slightly,  
the former supports individual connections, while the latter
relies on aggregation: by automatically determining the most
general rule matching a set of microflows to be handled in the same
manner. The latter 
can reduce the number of required
rules significantly in the fast path
and
the packets through the slow path.

\begin{figure}[t!]
	\centering
	\includegraphics[width=.9\columnwidth]{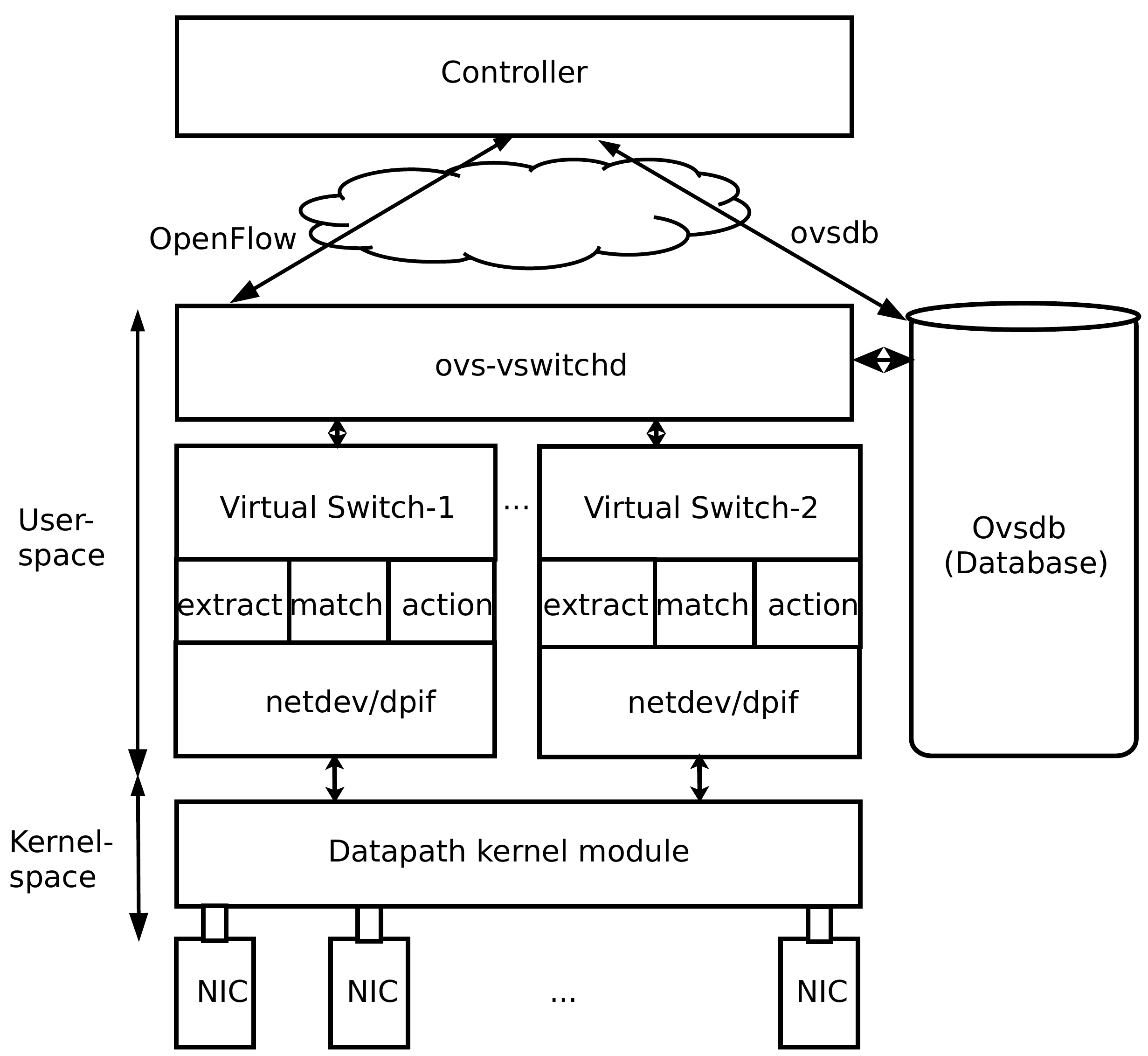}
	\caption{High-level architecture of Open vSwitch. Multiple virtual
switches interact with the datapath kernel module for packet processing and networking. The slow path is in user-space and the fast path is in kernel-space.
The virtual switches are instantiated by the \textit{ovs-vswitchd} deamon which
obtains configuration information from the ovsdb-server.
The controller manages and configures \textit{ovs-vswitchd} and \textit{ovsdb}
using OpenFlow and resp. ovsdb protocols over the network.}
	\label{fig:softwareSwitch}
\end{figure}

A high-level overview of the architecture of OvS 
is show in Fig.~\ref{fig:softwareSwitch}. 
OvS comprises of an \textit{ovsdb} database that stores relevant
switch configuration information such as switch/bridge name,
associated ports, match/action rules, port speeds, etc.
Necessary bridges/switches, ports, etc. are instantiated
using the configuration from the database by \emph{ovs-vswitchd}.
The database can be modified by the controller using the ovsdb protocol.
\emph{ovs-vswitchd} also manages the datapath kernel module.
The three stage packet processing pipeline is depicted
by the extract, match, and action.
The datapath kernel
module interfaces with user-space using
a modular datapath interface.
\emph{ovs-vswitchd} is managed
by the controller using OpenFlow.


\section{Compromising Cloud Systems\\via the Data Plane}
\label{sec:compromise}
This section presents a first security analysis of
virtual switches. In particular, we identify 
and characterize 
properties of  
virtual switches which may be exploited for attacks.  
Accordingly, we also compile and present 
a threat model. 

\subsection{Characterizing Virtual Switch Vulnerabilities}

We identify the following properties which 
are fundamental for virtual switches.
As we will demonstrate, in combination, 
they introduce serious vulnerabilities which are cheap to exploit,
i.e., by an attacker with low resources:
\begin{enumerate}
\item \emph{Security Assumptions:} 
Virtual switches often run with elevated (root) privileges by design.

\item \emph{Virtualized Data Plane:} Virtual switches reside
in virtualized servers (\emph{Dom0}), and are hence co-located with other, 
possibly critical, cloud software, including controller software.

\item \emph{Logical Centralization and Bidirectional Communication:} 
The control over programmable data plane elements 
is often outsourced and consolidated to 
a logically centralized software. For communication
between controller(s) and data plane elements,
bidirectional communication channels are used.  

\item \emph{Support for extended protocol parsers:} 
It is tempting to exploit 
the flexibilities of programmable virtual switches
to realise functionality which goes beyond the
basic protocol locations of normal switches, e.g., 
trying to parse the transport protocols (e.g., TCP) in switched packets
or handling protocols such as MPLS in a non-standard manner.
\end{enumerate}

In combination, these properties can render data plane attacks harmful: 
a software vulnerability in the packet processing logic
of a virtual swich running with root
privileges can be exploited
to not only compromise the virtual switch, but also
the underlying host operating system.
Hence co-located applications and tenants
are also compromised (e.g., an attacker can extract private keys,
monitor network traffic, etc.).
From there,
the controller(s) can be compromised. The
attacker can leverage the logically centralized
view to manipulate the flow
rules, possibly violating essential network security policies, or to
gain access to other resources in the cloud: For example,
the attacker may modify  
the identity management service (e.g., Keystone)
or the images (e.g., to install backdoors) which
are used to boot tenant VMs.

\subsection{Threat Model: Virtual Switch}

The vulnerabilities characterized above suggest that 
the data plane should not be considered trustworthy
and may not be treated as a black box.
It also highlights
that even an unsophisticated attacker
with very limited resources can cause significant harm,
far beyond 
compromising a single vulnerable switch.

Accordingly, we now introduce our threat model.
The attacker's target environment in this model is 
a cloud infrastructure that utilizes
virtual switches for network virtualization.
The cloud is hosted in a physically secured facility i.e., access to the facility is restricted.
Its services are either public, private or a hybrid.
If the cloud is not public, we assume that the attacker is a malicious insider.
We assumme that the cloud provider may follow a security best-practises guide~\cite{openstackSecurityGuide}: It may therefore 
create three or more isolated networks (physical/virtual) 
dedicated towards management, tenants/guests and external traffic.
Furthermore, we assume that the same virtual switches such as OvS 
are used across all the servers in the cloud.

The attacker is financially limited and initially has access to limited resources in the cloud (i.e the resources of a VM). Additionally, the attacker controls a computer that is reachable from the cloud under attack.
After compromising the cloud, the attacker can have control over the cloud resources:
it can perform arbitrary computation, create/store arbitrary data, and lastly transmit arbitrary data to the network.

\section{Proof-of-Concept: A Case Study\\of OvS and OpenStack}\label{sec:proofofconcept}

To demonstrate the severity of virtual switch attacks,
we present proof-of-concept attacks with OvS in OpenStack.
OvS is a
popular and widely-deployed state-of-the-art virtual switch
(default virtual switch in OpenStack),
supporting logically centralized control and OpenFlow. 
Moreover, the OvS daemon (ovs-vswitch) executes with root privileges
(recall the virtual switches properties in Section~\ref{sec:compromise}).

\subsection{Bug Hunting Methodology}

We use a simple coverage-guided fuzz testing 
to elicit crashes in the packet parsing subsystem of OvS.
The reason we chose this subsystem is due to the fact
that it directly accepts input (packets) from the attacker.
In fact, to find the vulnerabilities presented in this paper, 
it was sufficient to fuzz only a small fraction (less than 2\%) of the total 
executions paths of \emph{ovs-vswitchd}.

In our methodology, all crashes reported by the fuzzer were triaged to ascertain 
their root cause.
The test harness ($test-flows$) accepts two user inputs, namely, the 
flow configuration file ($flows$), and an incoming network 
packet ($pkt$) to be processed by the switch.
The configuration takes the form of flow rules: 
the list of match/action rule statements that fully
determine the switch's state machine.
During the switch's operation, an incoming packet is 
parsed and matched against flow rules.
A majority of our effort was focussed on fuzzing the
 \textit{flow extraction} code---the OvS subsystem that parses incoming packets.
For our tests, we used the American Fuzzy Lop (AFL) open-source 
fuzzer version 2.03b and OvS source code (v2.3.2, v2.4.0 and v2.5.0) 
recompiled with AFL instrumentation.

\subsection{Identified Vulnerabilities}

We discovered three unique vulnerabilities:

\begin{itemize}
\item Two stack buffer overflows in MPLS 
parsing code in OvS (2.3.2, 2.4.0): The stack buffer overflows 
occur when a large MPLS label stack packet exceeding a pre-defined 
threshold is parsed (2.3.2),
when an early terminating MPLS label packet is parsed (2.4.0).
\item An integer underflow 
which leads to a heap buffer overread in the IP packet parsing code 
in OvS (2.5.0): 
The underflow and subsequent overread occurs 
when parsing an 
IP packet 
with zero total length or a total length lesser than the 
IP header length field.
\end{itemize}

\begin{figure}[t!]
	\centering
    \includegraphics[trim=2.5cm 26.5cm 9.75cm 0.0cm,clip=true,width=.9\columnwidth]{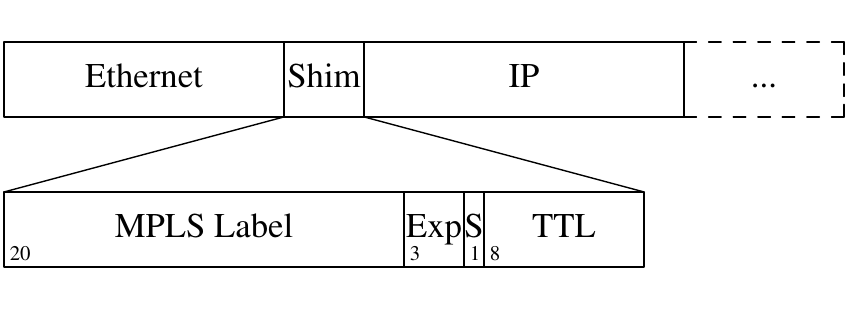}
    \caption{The Shim header is placed between the Ethernet and IP headers.
The shim header (MPLS label) is a 32 bit value that includes a label
used to make forwarding decisions.
The Exp field is 3 reserved bits.
If set to 1, the S field indicates that the label is the bottom (end) of the label stack.
The 8 bit TTL field indicates the Time-To-Live.
}
    \label{fig:shimPacket}
\end{figure}

The fact that two vulnerabilities are related
to MPLS should not be too surprising: 
they relate to the fundamental
properties of virtual switches discussed in our security
analysis in the previous section.
Before delving into the details however,
in order to understand these attacks, we quickly
review
the MPLS label stack encoding (RFC 3032)~\cite{rfc3032}.
In Ethernet and IP based networks,
MPLS labels are typically placed between the 
Ethernet header (L2) and the IP header (L3),
in a so-called \emph{shim header}.
Multiple labels can be stacked: \emph{push} and
\emph{pop} operations are used to add resp.~remove
labels from the stack. 
Fig.~\ref{fig:shimPacket}
shows the position of the shim header/MPLS label 
and the structure as per RFC 3032.
The MPLS label is 20 bits long used to make
forwarding decisions instead of the IP address.
The Exp field is 3 bits of reserved space.
If set to 1, the S field indicates that the label
is the bottom of the label stack. It is a
critical piece of ``control'' information that determines
how an MPLS node parses a packet.
The TTL field indicates the Time-To-Live of the label.

With the MPLS label stack encoding in mind,
we now explain the buffer overflow vulnerabilities.
In the OvS 2.3.2 buffer overflow, the S bit
was not set for the entire label stack of 375 labels.
(375 labels for a 1500 Max.~Transmission Unit size).
In the OvS 2.5.0 buffer overflow, the label itself was
malformed i.e., it was less than 4 bytes.


\subsection{Weaponzing the Vulnerabilities}

To illustrate the potential damage 
and consequences
of these vulnerabilities,
we developed real-world exploits that leverage the 
discovered vulnerabilities.
Our exploits, at their core,
consist of simply sending a malformed packet to 
a virtual switch.
They achieve one of the following:
gain arbitrary code execution on, 
bypass an access control list of,
or deny service to the virtual switch.
Our attacks demonstrate that even a weak attacker can inflict huge 
damage in the cloud, compromising the confidentiality, integrity, and 
availability of the servers in the cloud as well as its tenants.
In the following,
we formulate our attacks on OvS in an OpenStack
cloud setting, validate the attacks
and estimate the impact of the attacks in our setup.

\subsubsection{Rein Worm Attack}

We provide an overview of the Rein worm attack before describing the 
exploitation process in more detail.
The Rein worm exploits the stack buffer overflow 
vulnerability in OvS (2.3.2).
Mounting a Return-Orienting Programming (ROP)~\cite{roemer2012return} 
attack on the server running \textit{ovs-vswitchd} from the VM,
provides the capability to spawn a shell on that server.
The shell can be redirected over the network to the remote attacker.
The attacker controlled server can then propagate the same attack
to the controller and from there on to all the other servers.
The centralized architecture of OpenStack and SDN
requires the controller
to be reachable from all servers and resp.~data planes in the network.
This inherent property provides the
necessary connectivity for worm propagation.
Furthermore, the network isolation using
VLANs and/or tunnels (GRE, VXLAN, etc.) do not
affect the worm once the server is compromised.

\begin{figure*}[t!]
  \centering
  \begin{subfigure}{.49\linewidth}
  	\centering
    \includegraphics[trim=0.0cm 0.0cm 0.0cm 1.0cm,clip=true,width=.9\columnwidth]{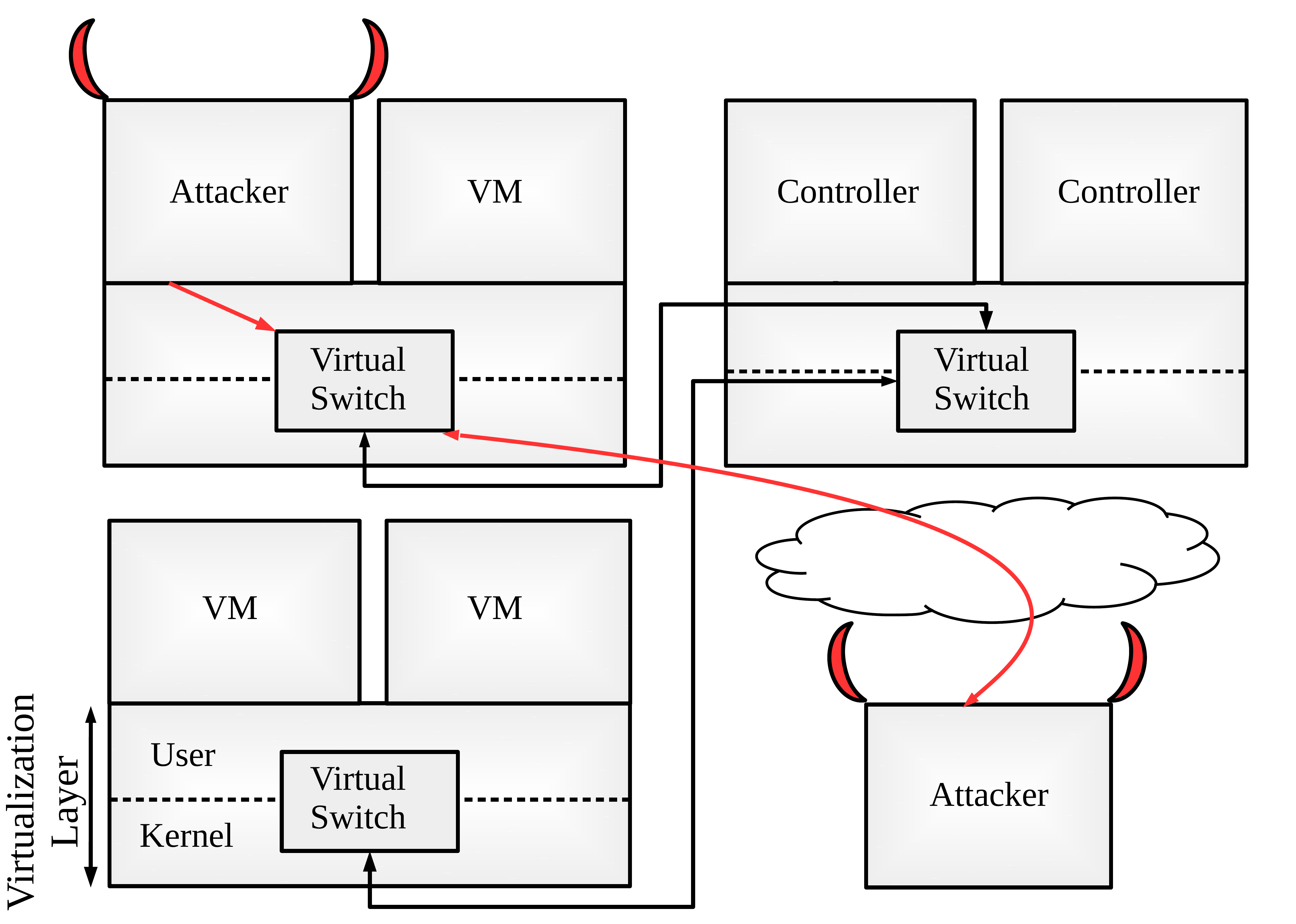}
		\caption{Step 1.}
\label{fig:longShimLabel}    
  \end{subfigure}\hfill
  \begin{subfigure}{.49\linewidth}
    \includegraphics[trim=0.0cm 0.0cm 0.0cm 1.0cm,clip=true,width=.9\columnwidth]{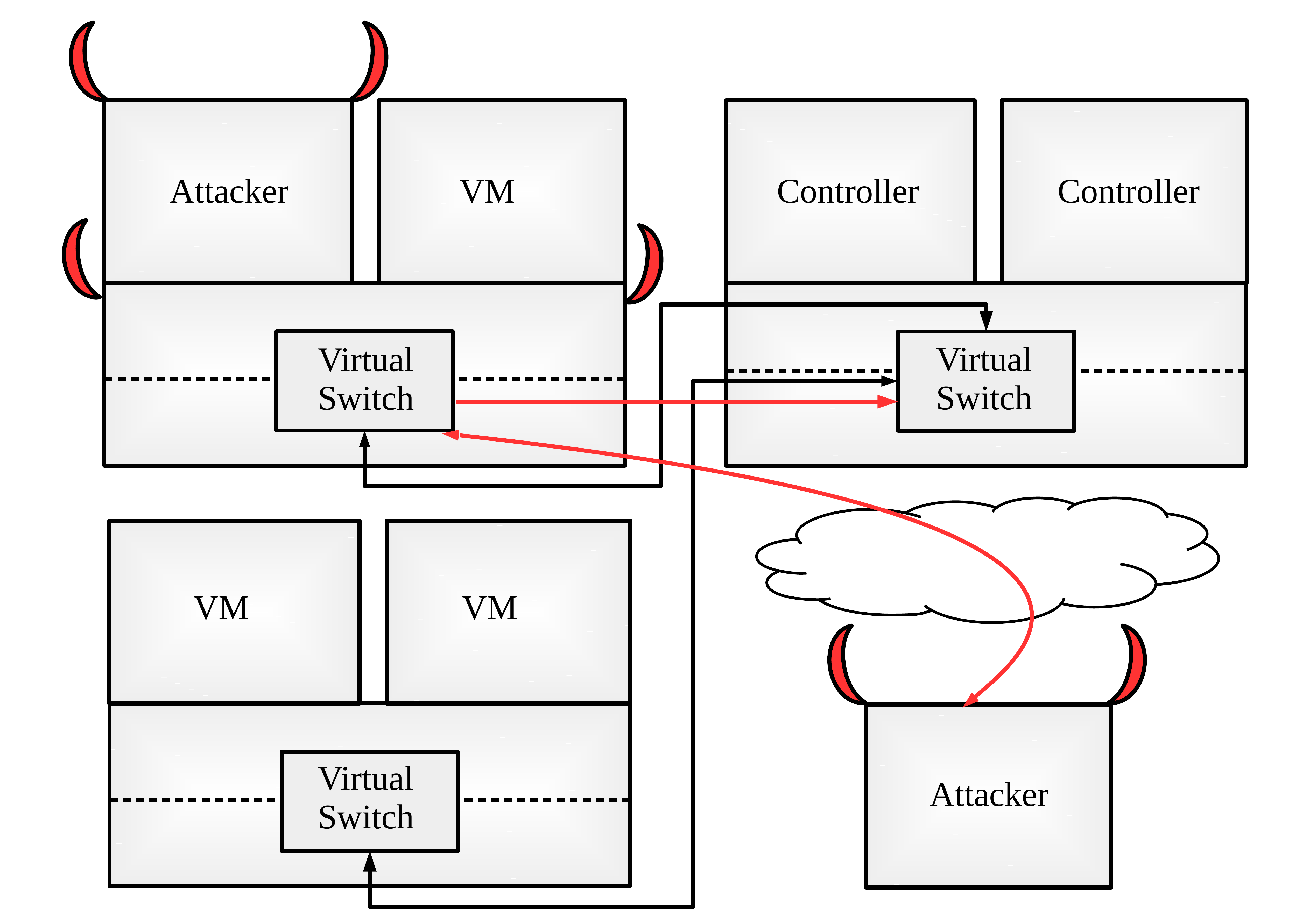}
		\caption{Step 2.}
  \label{fig:shortShimLabel}
  \end{subfigure}
  \begin{subfigure}{.49\linewidth}
    \includegraphics[trim=0.0cm 0.0cm 0.0cm 1.0cm,clip=true,width=.9\columnwidth]{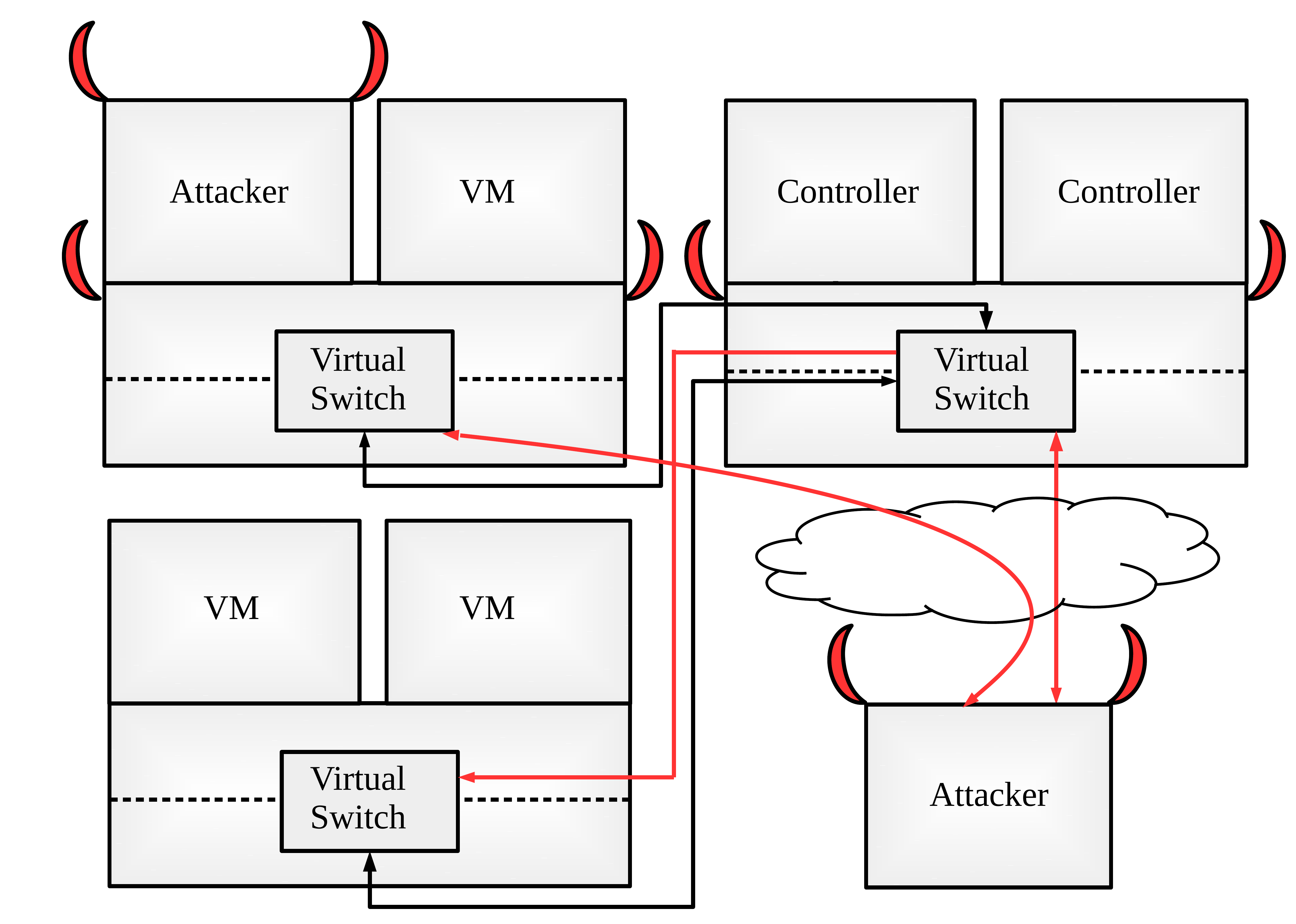}
		\caption{Step 3.}
  \label{fig:shortShimLabel}
  \end{subfigure}
  \begin{subfigure}{.49\linewidth}
    \includegraphics[trim=0.0cm 0.0cm 0.0cm 1.0cm,clip=true,width=.9\columnwidth]{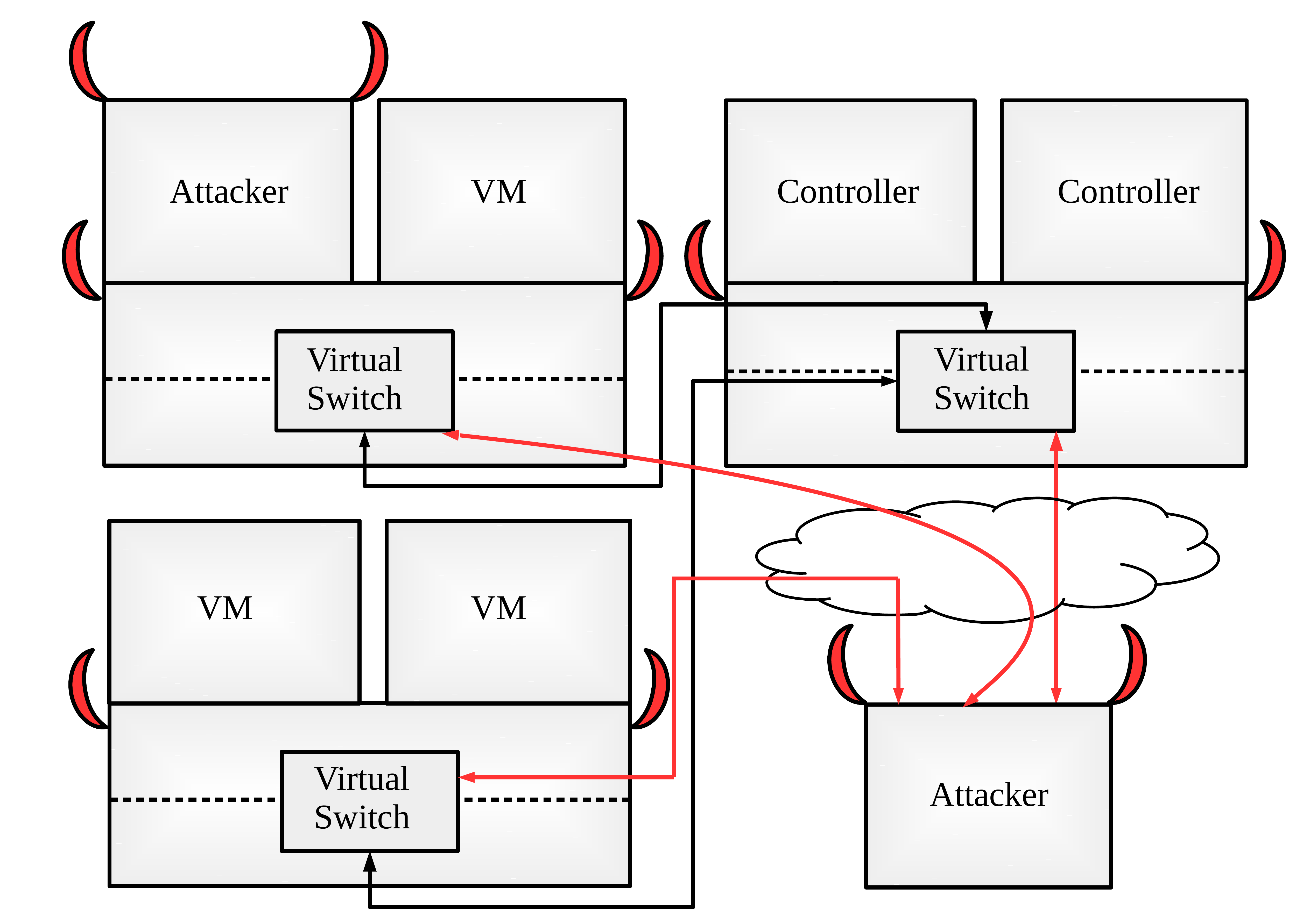}
		\caption{Step 4.}
  \label{fig:shortShimLabel}
  \end{subfigure}
  \caption{The steps that are involved in
the Rein worm. In step 1, the attacker VM sends
a malicious packet that compromises its server, giving
the remote attacker control of the server. In step 2, the attacker
controlled server compromises the controllers' server,
giving the remote attacker control of the controllers'
server. In step 3, the compromised controllers' server propagates
the worm
to the remaining uncompromised server.
Finally in step 4, all the servers are controlled
by the remote attacker.}
  \label{fig:wormAttackSteps}
\end{figure*}

Fig~\ref{fig:wormAttackSteps} visualizes the steps of the Rein worm.
In step 1, the Rein worm originates from an attacker-controlled (guest) VM 
within the cloud. It can compromise the host operating system (OS) of the server
due to the exploitable virtual switch.
With the server under the control of the remote attacker,
in step 2, the worm then propagates to the controller's server
and compromises it. With the controller's server
also under the control of the remote attacker, the worm
moves toward the remaining uncompromised server(s).
Finally, in step 4, all the servers are under the control
of the remote attacker.


\begin{figure}[t!]
    \center
	\includegraphics[width=.6\columnwidth]{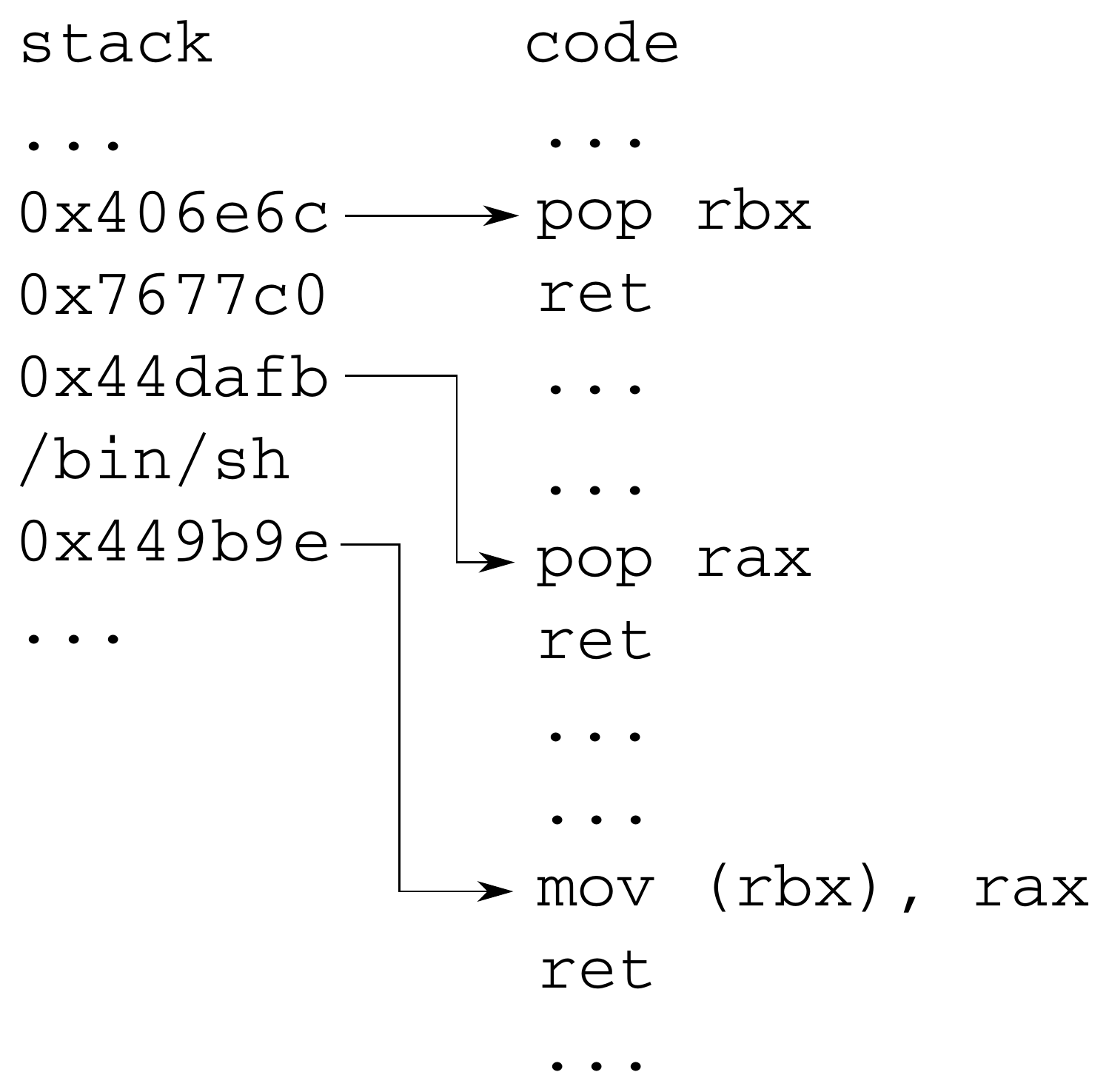}
	\caption{ROP chain that writes '/bin/sh' into
memory location '0x7677c0'.}
	\label{fig:rop}
\end{figure}

\noindent \textit{Exploit.}
A ROP attack places addresses of reused code snippets 
(gadgets) from the vulnerable program on the stack.
A gadget typically consists of one or more operations followed 
by a return.
After executing each gadget, the return will pop the 
address of the next gadget into the instruction pointer.
Figure~\ref{fig:rop} shows an example of writing the value 
'/bin/sh' into memory location '0x7677c0' using ROP.

A constraint for the ROP payload
is that the gadget addresses have to have their 
16th bit unset, i.e., the S bit in the MPLS label is zero.
We modified an existing ROP payload generation tool 
called Ropgadget~\cite{gadgettool} to meet this constraint.
To give the attacker a shell prompt at an IP 
address of choice, the following command
was encoded into the ROP payload:
\parbox{\textwidth}{\small\texttt{bash -c "bash -i $>$\& /dev/tcp/<IP>/8080 0$>$\&1"}}

\noindent \textit{Worm.}
There are multiple steps involed in propagating 
the worm.
All steps are deterministic and hence 
scriptable.
To propagate the Rein worm in our
test environment, we wrote a shell script.
The main steps that are:
\begin{enumerate}
\item Install a patched \textit{ovs-vswitchd} on the compromised host.
This is required to have OvS forward the attack payload from the
compromised server.
\item Obtain the exploit payload from an external location 
(public HTTP) if necessary.
\item Identify the correct network interface to the controller.
\end{enumerate}

\subsubsection{Long Shim Attack and Short Shim Attack}
\begin{figure}[t!]
	\centering
	\includegraphics[trim=0.0cm 0.0cm 0.0cm 0.0cm,clip=true,width=.9\columnwidth]{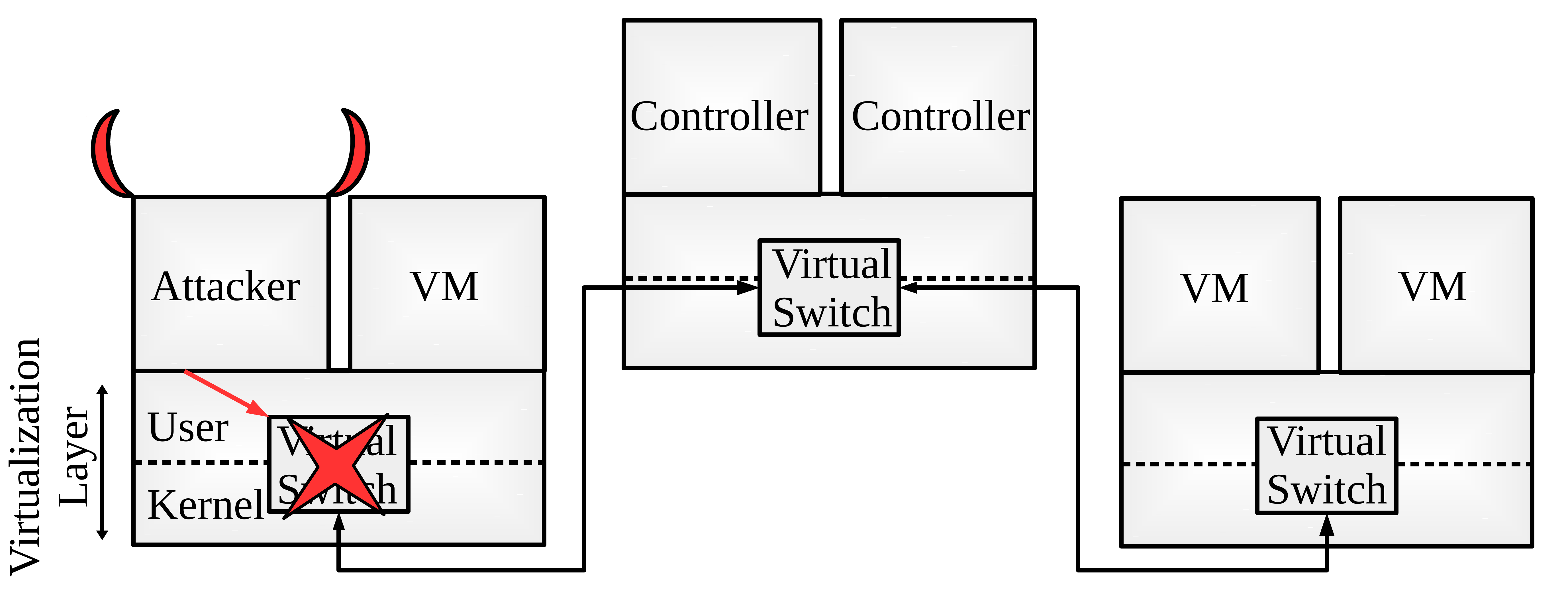}
	\caption{An attacker-controlled VM attacks the virtual
switch of its server using the Short Shim Attack or Long Shim Attack packet, disrupting
network connectivity for the server.}
	\label{fig:ShimAttack}
\end{figure}

\begin{figure*}[t!]
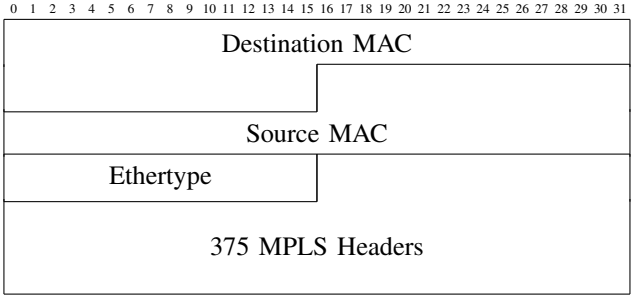
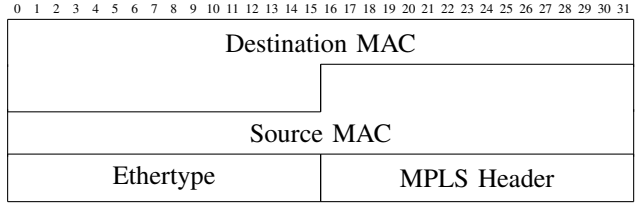

  \centering
  \begin{subfigure}{.49\linewidth}
		\begin{bytefield}[bitwidth=0.74em]{32}
		\bitheader{0-31} \\
		\wordbox[lrt]{1}{Destination MAC} \\
		\bitbox[lrb]{16}{} & \bitbox[lrt]{16}{}\\
		\wordbox[lr]{1}{Source MAC} \\
		\bitbox[lrbt]{16}{Ethertype} & \bitbox[lrt]{16}{}\\
		\wordbox[lrb]{2}{375 MPLS Headers} \\
		\end{bytefield}
		\caption{Long Shim Attack packet format.}
\label{fig:longShimLabel}    
  \end{subfigure}\hfill
  \begin{subfigure}{.49\linewidth}
		\begin{bytefield}[bitwidth=0.74em]{32}
		\bitheader{0-31} \\
		\wordbox[lrt]{1}{Destination MAC} \\
		\bitbox[lrb]{16}{} & \bitbox[lrt]{16}{}\\
		\wordbox[lr]{1}{Source MAC} \\
		\bitbox[lrbt]{16}{Ethertype} & \bitbox[lrbt]{16}{MPLS Header}\\
		\end{bytefield}
                \vspace{3.5em}
		\caption{Short Shim Attack packet format.}
  \label{fig:shortShimLabel}
  \end{subfigure}
  \caption{The Layer 2 Ethernet frame starts with the 
  Destination MAC address,
followed by the Source MAC address and then the Ethertype. 
The Ethertype value for MPLS unicast packets is 0x8847. 
The Long Shim Attack packet is malformed since 1500 bytes of data are 
filled with MPLS headers. The MPLS headers encapsulate the ROP 
payload in the Rein worm. The Short Shim Attack packet is malformed as 
the label is only 16 bits long.
Note that the Preamble, Frame Delimiter and Frame Check 
Sequence fields from the Ethernet frame are not shown for the sake of simplicity.}
  \label{fig:shimLabels}
\end{figure*}

The Long Shim Attack and the Short Shim Attack are Denial of Service 
(DoS) attacks targeted at \textit{ovs-vswitchd} (2.3.2 and 2.4.0).
They leverage stack buffer overflows to crash 
the daemon, thereby denying network
service to the host and guest on the server.
Figure~\ref{fig:ShimAttack} visualizes the attack.
To launch a DoS attack, an attacker simply needs to
send the malformed packet out its VM(s).
The attack causes a temporary network outage:
for
all guest VMs that connect to the virtual switch,
for the host to connect to the controller
and also for other servers in the cloud.
Repeated attacks increase the longevity of the outage.
Figure~\ref{fig:shimLabels} shows the attack packet for the Long Shim Attack and the Short Shim Attack.
The only difference in the structure of the packets used in the 
two attacks is that one contains an oversized Shim header (MPLS label stack)
while the other contains an undersized (less than four bytes) 
Shim header.

\subsubsection{Access Control List Bypass}
\label{sec:ovs250}

The Access Control List (ACL) bypass leverages a 2-byte heap 
buffer overread in the IP packet parsing code of \textit{ovs-vswitchd}.
The heap overread vulnerability is caused by the unsanitized use 
of the total length field present in the IP header.
The vulnerability results in the IP payload (e.g., TCP, UDP) being 
parsed for packets with an invalid IP header, ultimately 
resulting in an ACL bypass.
In other words, packets that should have been dropped at the
switch are instead forwarded to the next hop in the network.
In addition, if the malformed IP packets were to reach a 
router from OvS, it may elicit ICMP error messages 
from the router as per RFC 1812~\cite{rfc1812} causing unnecessary 
control plane traffic at the router and OvS.
However, end hosts are not affected by this vulnerability 
since most OS kernels are expected to drop such packets.

\subsection{Attack(s) Validation and Impact}

We used Mirantis 8.0 distribution of OpenStack 
to create our test setup and validated
the attacks.
The test setup consists of a server (the fuel master node) 
that configures and deploys other OpenStack nodes (servers) such as the controller, 
compute, storage, network, etc.
Due to our limited resources,
we created one controller and one compute node in
addition to the fuel master node using the
default configuration Mirantis 8.0 offers.

Using our setup, we deployed the Rein worm and  measured the time 
it takes for an attacker to obtain root shells on the compute and 
controller nodes originating from a guest VM on the compute node.
We started the clock from the time the Rein worm was sent from the 
VM and stopped the clock when a root shell was obtained 
on the controller node.
We found that in our environment it took 21s which involved 
12s of sleep time
(for restarting \textit{ovs-vswitchd} and \textit{neutron-agent}
on the compute node) 
and 3s of download time (for the patched \textit{ovs-vswitchd},
shell script, and exploit payload).
To restore network services on the controller node, 
a sleep time of 60s was required.
From this we can extrapolate that compromising
100 compute nodes and 1 controller node,
would take less than 100s,
assuming that from the controller node, the Rein worm 
can reach all other nodes at the same time.
Deploying the Short Shim Attack and Short Shim Attack attacks in our
setup,
we can create an outage time,
in the order of 4-5s. Obviously,
depending on the configuration of the
virtual switch and computing power of
the node, the outage time may vary.


\subsection{Summary}

The OvS and OpenStack case study provides
a concrete instance of the
theoretical threat model derived in Section~\ref{sec:compromise}.
Indeed, we have demonstrated that the NIC, the fast-path, and the 
slow-path of OvS are all
facing the attacker. In particular, the attack can leverage 
(1) existing security assumptions (OvS executes with
root privileges), (2) virtualization (collocation with
other critical cloud applications), (3) logical centralization 
(bidirectional channels to the controller),
as well as non-standard MPLS parsing,
to launch a potentially very harmful attack. 
This is worrisome, and raises the question of possible
countermeasures: the subject of the next section.

\section{Countermeasure Evaluation}
\label{sec:eval}
Mitigations against attacks such as the ones we were able to 
perform against
OvS, have been investigated intensively in the past.
Proposals such as
MemGuard~\cite{Cowan98}, control flow integrity~\cite{Abadi05}
and position independent executables
(PIEs)~\cite{payer2012} could have prevented our attacks. 
Newer approaches, like Safe (shadow) Stack~\cite{186159} can even prevent 
ROP attacks. By separating the safe stack (return addresses, code pointers) 
from the unsafe stack (arrays, variables, etc.), control flow integrity
can be preserved, while data-only attacks may remain possible~\cite{chen2005}.
The downside of these mitigations is their potential performance overhead.
MemGuard imposes a performance overhead of 3.5--10\%~\cite{Cowan98}, while
PIEs have a performance impact of 3--26\%~\cite{payer2012}.

Performance evaluation of these mitigations in prior work~\cite{Cowan98, payer2012, 186159}
naturally focused on the overall system performance and binary size with
applied mitigations. As Table~\ref{tab:binarySize} shows, the available
mitigations do indeed increase the size of the \textit{ovs-vswitchd} and \textit{openvswitch.ko} binaries
significantly. However, OvS performance mainly depends on two 
metrics: forwarding latency and forwarding throughput. To determine the
practical impact of available and applicable mitigation, we hence designed a
set of experiments that evaluate the relevant performance impact for OvS
forwarding latency and performance.

\begin{figure}[t!]
	\includegraphics[width=.9\columnwidth]{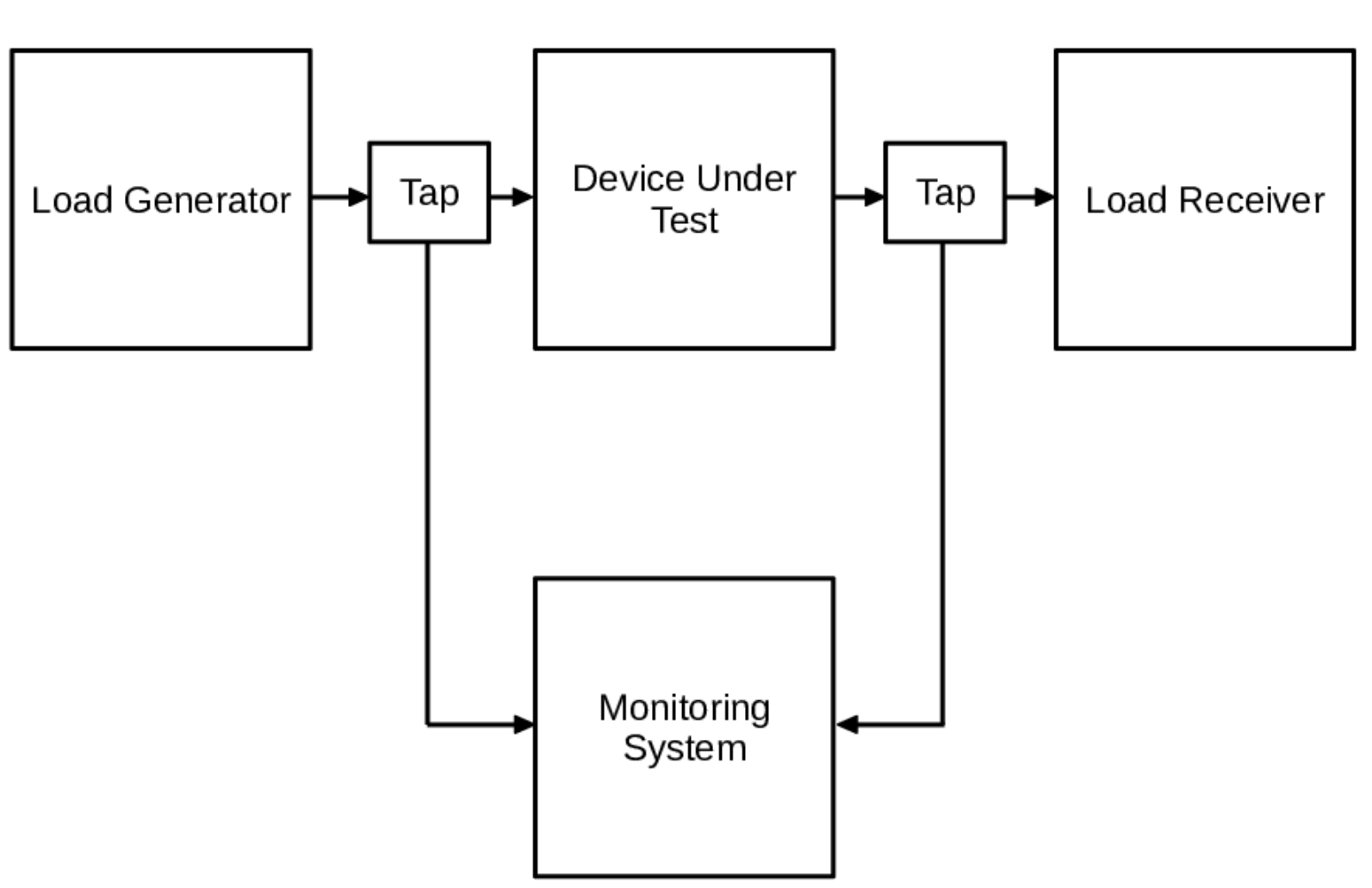}
	\caption{Setup of the performance evaluation.}
	\label{fig:perfTestbed}
\end{figure}

\begin{table}[t]
\centering
\begin{tabular}{c c c}
\toprule
	Binary type&Binary size(MB) & \% of Baseline\\ \hline
\midrule
	ovs-vswitchd baseline&1.84 & \\
	ovs-vswitchd with stack protector and pie&2.09 & +13.59\% \\
	openvswitch.ko baseline&0.16 & \\
	openvswitch.ko with grsecurity&0.21 & +31.25\% \\
\bottomrule
\end{tabular}
\caption{Size comparison of \textit{ovs-vswitchd} and \textit{openvswitch.ko} binaries using \emph{gcc} countermeasures and grsecurity patch respectively.}
\label{tab:binarySize}
\end{table}

\noindent \textbf{Evaluation Parameters:} We evaluate forwading latency and
throughput in eight different common cases. We compare a vanilla Linux kernel
(v4.6.5) with the same kernel integrated with $\grsec$ patches (v3.1), which, e.g.,
protects kernel stack overflows, address space layout randomization, ROP defense, etc. For both kernels, we evaluate
OvS-2.3.2, once compiled with {\tt -fstack-protector-all} for unconditional
stack canaries and {{\tt -fPIE} for position independent executables, and once
compiled without these two features. As \emph{gcc}, the default compiler for the Linux
kernel, does not support the feature of two seperate stacks (safe and unsafe)
we did not evaluate this feature, even though 
it would be available with \emph{clang}
starting with version~3.8. In addition, to compile-time security options we
also evaluate the impact of traffic flowing either exclusively through the fast
or slow path. For the slow path exclusive experiments we disabled a default
configuration option \emph{megaflows cache}. This disables
generic fast path matching rules
(Sec.~\ref{sec:virtualSwitches}),
following current best practices for benchmarking OvS, see 
Pfaff et al.~\cite{pfaff2015design}.

\noindent \textbf{Evaluation Setup:} For our measurements, 
we utilized four
systems, all running Linux kernel (v4.6.5) compiled with gcc (v4.8). The
systems have 16GB RAM, two dual-core AMD x86/64 2.5GHz and four Intel Gigabit
NICs. The systems are interconnected as illustrated in
Figure~\ref{fig:perfTestbed}. One system serves as the Load Generator (LG)
connected to the Device Under Test (DUT), configured according to the different
evaluation parameters. The data is then forwarded by OvS on the DUT to a
third system, the Load Receiver (LR). The connections between LG and DUT and LR
and DUT respectively are run through a passive taping device. Both taps are
connected to the fourth system. Data collection prior and post the DUT is done
on one system to reduce the possible impact of clock-scew. Given the small values
we intend to measure, we acknowledge that some timing noise may occur.
To counteract that, we selected appropriately large sample sizes.

\noindent \textbf{Throughput Evaluation:} For the throughput evaluation we
created files containing a constant stream of 60 byte UDP packets.  We opted for
60 byte packets in order to focus on the packets per second (pps) throughput
instead of the bytes per second throughput, as pps throughput indicates performance
bottlenecks earlier~\cite{jacobson1988congestion}. These were then replayed from the
LG via the DUT to the LR using \emph{tcpreplay}. Each experimental run consists of 120
seconds where we replay at rates between 10$k$ and 100$k$ packets per second,
incremented in steps of 10$k$ pps. For the all-slow-path experiments, each of the
generated packets used a random source MAC address, as well as source and
destination IPv4 address and random source and destination port. For the
all-fast-path experiments we re-sent packets with the same characteristics
(source, destination, etc.) that match a pre-installed flow rule.

\begin{figure}[tb!]
	\centering
	\begin{subfigure}{.99\linewidth}
		\includegraphics[trim=1.0cm 2.3cm 2.1cm 0.0cm,clip=true,width=.99\columnwidth]{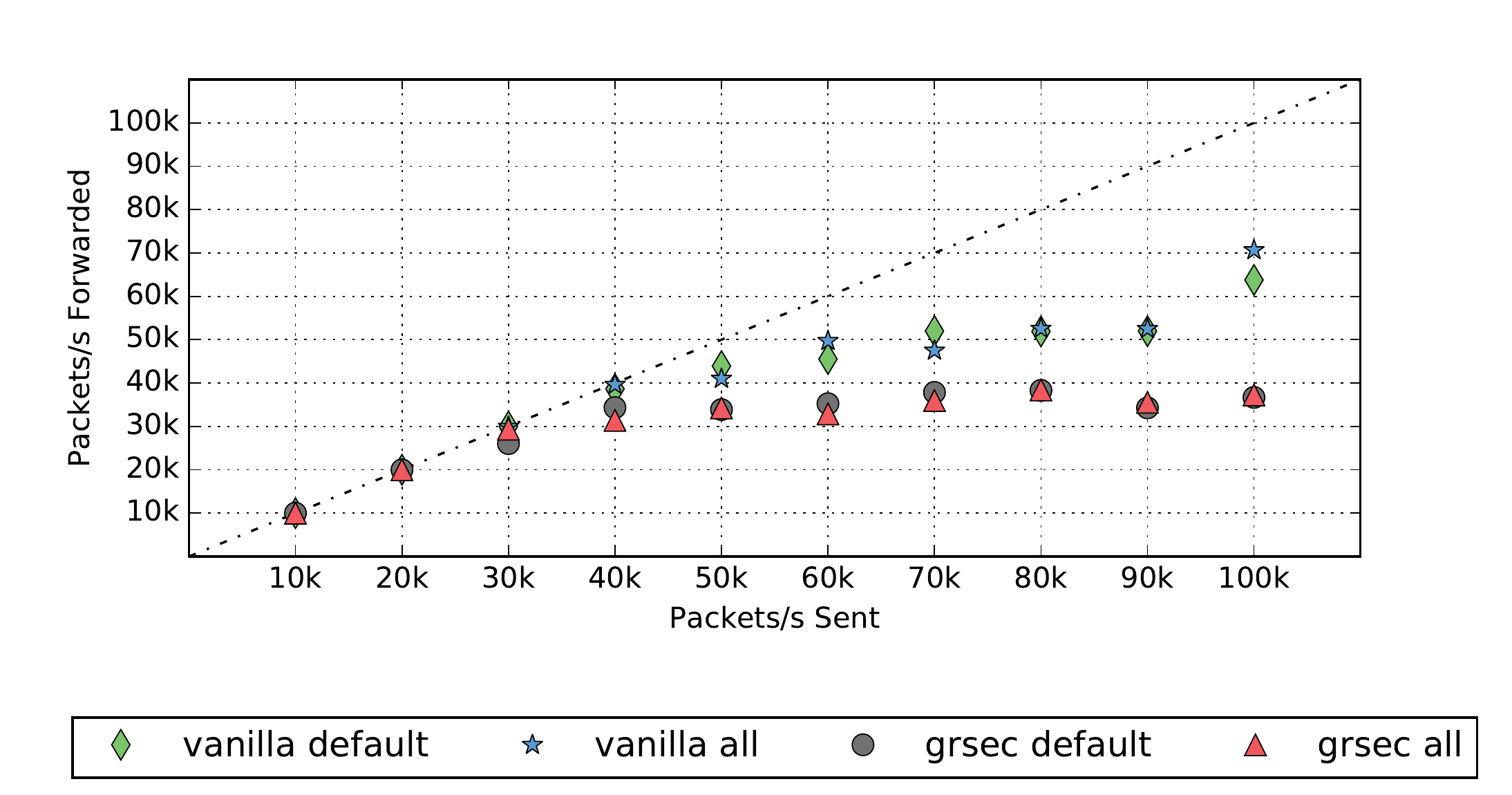}
	\end{subfigure}
	\\
	\begin{subfigure}{.99\linewidth}
		\includegraphics[trim=1.0cm 0.0cm 0.5cm 10.0cm,clip=true,width=.99\columnwidth]{plot_tp_0fp_100sp}
	\end{subfigure}

	\caption{Slow path throughput measurements for OvS compiled with gcc with and without countermeasures on a vanilla kernel and a grsecurity enabled kernel.}
	\vspace{-2mm}
	\label{fig:throughput}
\end{figure}

\begin{figure*}[tb!]
	\centering
	\begin{subfigure}{.49\linewidth}
		\centering
		\includegraphics[trim=0.0cm 0.0cm 2.0cm 0.0cm,clip=true,width=.99\columnwidth]{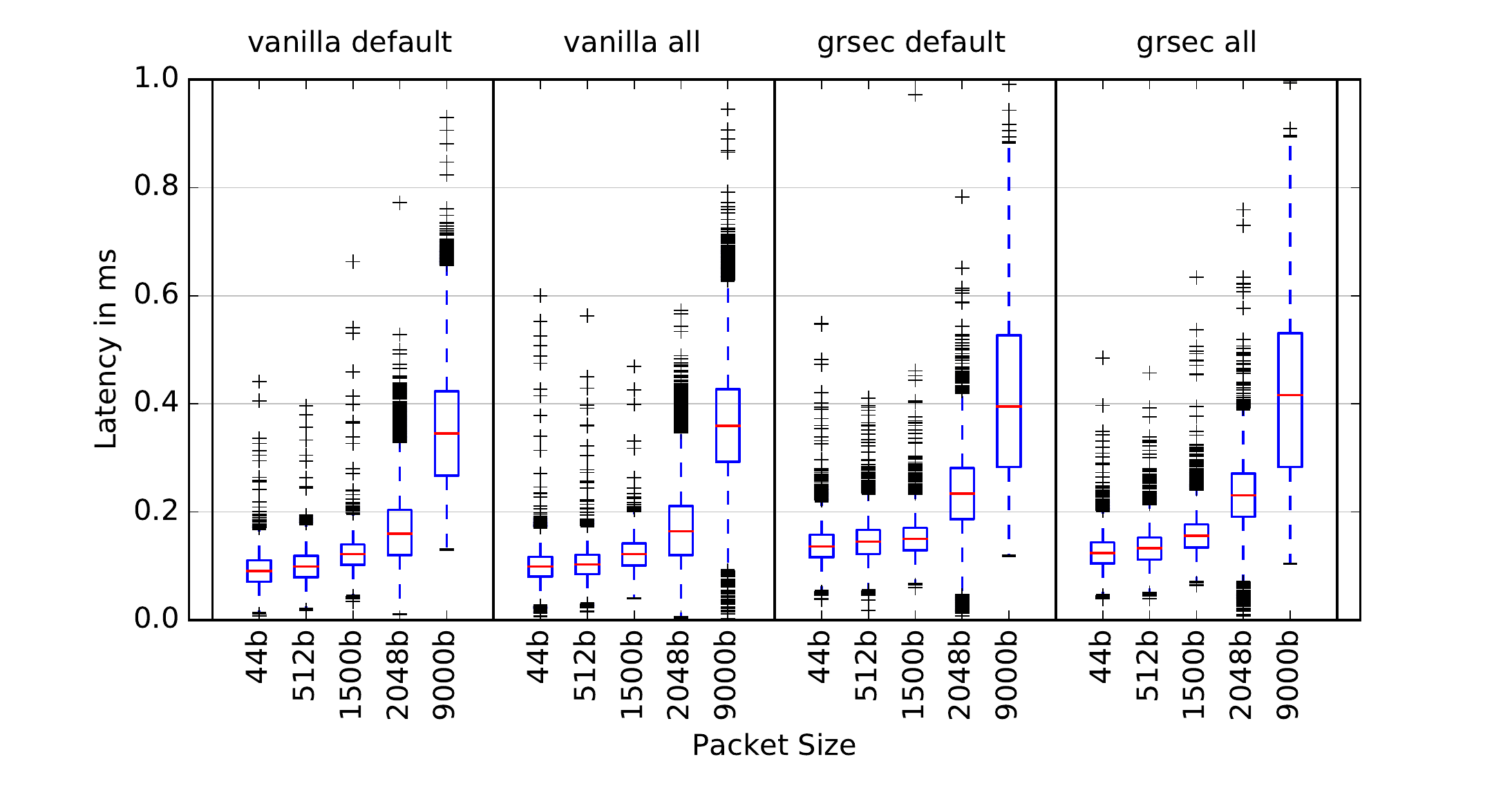}
		\caption{0\% fast path, 100\% slow path}
		\label{fig:latency-100s}
	\end{subfigure}\quad
	\begin{subfigure}{.49\linewidth}
		\centering
		\includegraphics[trim=0.0cm 0.0cm 2.0cm 0.0cm,clip=true,width=.99\columnwidth]{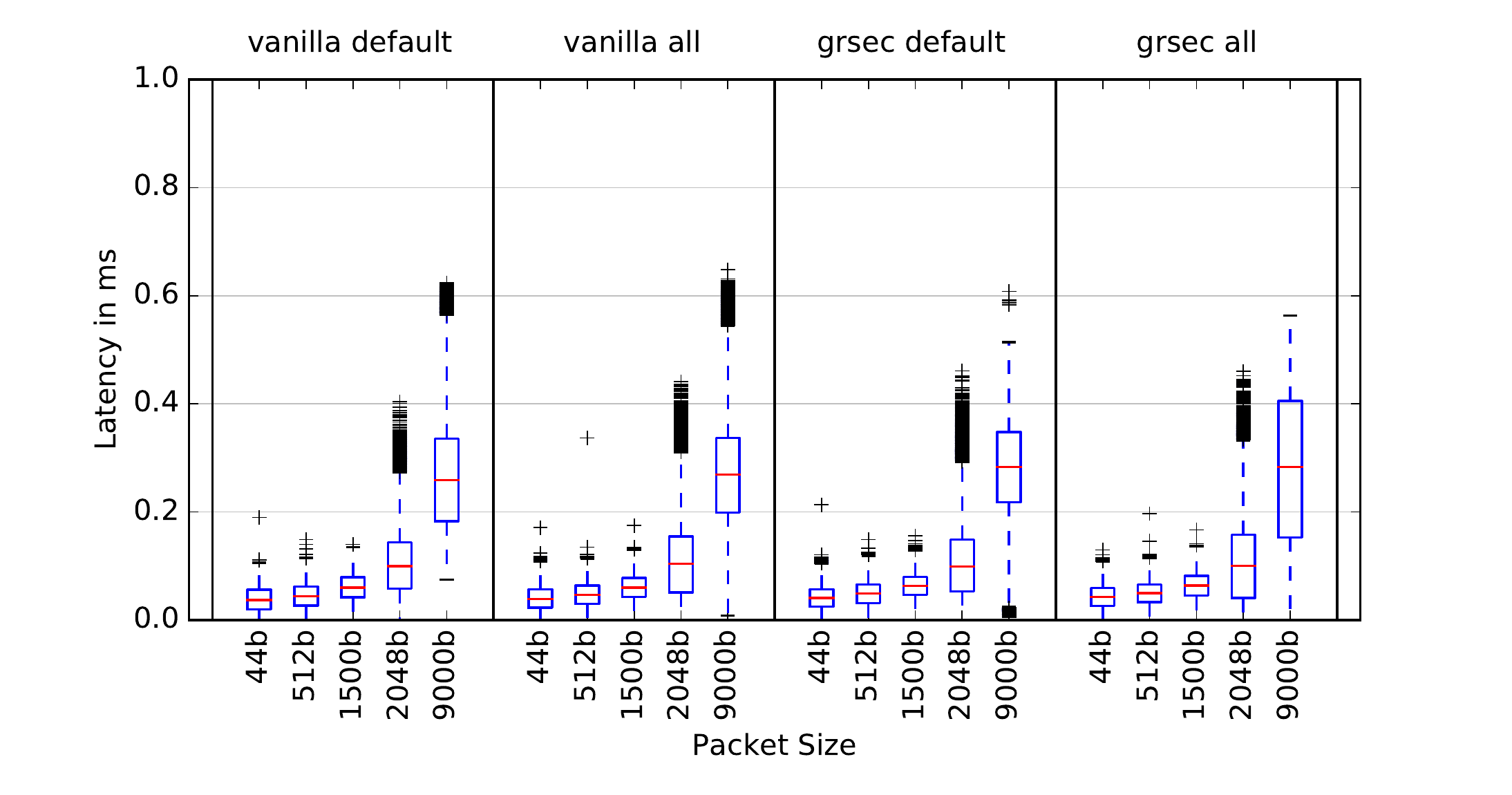}
		\caption{100\% fast path, 0\% slow path}
		\label{fig:latency-100f}
	\end{subfigure}
\caption{Latency measurements for OvS compiled with gcc with and without countermeasures on a vanilla kernel and a grsecurity enabled kernel.}
\label{fig:thorughput}
\end{figure*}

An overview of the results for the slow path throughput measurements are
depicted in Figure~\ref{fig:throughput}.  Packet loss for the vanilla kernel
first lies above 50$k$ pps, while the experiments on the \grsec{}
enabled kernel already exhibit packet loss at 30$k$ pps. Apart from the \grsec{}
kernel patches, we do not observe a significant impact of userland security
features on the forwarding performance of OvS. The results for the fast path
measurements are not illustrated, as we observed an almost linear increase with no
impact of the chosen evaluation parameters at all\footnote{We do note that
this is different if the \emph{megaflows} feature is disabled for fast path measurements. In that
case we observe a similar curve as in Figure~\ref{fig:throughput}, with
packet loss first occuring around 10$k$ later and a higher difference around
20$k$ between the asymptotic development of the values between \grsec{} and
vanilla. However, to remain compatible with the best current practices in
existing literature and as this work's main focus is not the fullscale
performance evaluation of OvS, we adhere to the approach by Pfaff et
al.~\cite{pfaff2015design}}. An impact of the parameters may exist with higher
pps counts. However, our load generation systems were unable to provide
sufficiently stable input rates beyond 100$k$ pps.

\noindent \textbf{Latency Evaluation:}  For the latency evaluation, we studied
the impact of packet size on OvS forwarding. From the possible packet sizes we select 44$b$
(minimum packet size), 512$b$ (average packet), and 1500$b$ (Maximum Maximum
Transmission Unit (MTU)) packets from the legacy MTU range; in addition, we also
select 2048$b$ packets as small jumbo frame packets, as well as 9000$b$ as
maximum jumbo frame sized packets. For each experimental run, i.e., packet size
for one of the parameter sets, we continuously send 10,500 packets from LG to LR
via the DUT with a 100ms interval. The packet characteristics correspond to
those from the throughput evaluation, i.e., random packets for the slow path
and repetitive packets matching a rule for the fast path. To eliminate possible
build-up or pre-caching effects, we only evaluate the later 10,000 packets for
each run.

The results for the latency evaluation are depcited in
Figure~\ref{fig:latency-100s} for the slow path and
Figure~\ref{fig:latency-100f} for the fast path experiments respectively.  For
the slow path, we see that \grsec{} (grsec default and grsec all)
imposes an overall increase in latency of approximately 25--50\%, depending on the packet size.
This increase is even higher for jumbo frames.  At the same time, also the
variance of values is increased by the use of \grsec{}. Still, we cannot
observe any significant impact of the userland protection mechanisms for
slow path latency for neither a vanilla nor a \grsec{} enabled kernel. These
observations also support the findings from the throughput measurements depicted
in Figure~\ref{fig:throughput}, where we also observe consistently lower performance
of approximately 25\%
for \grsec{} enabled systems on the slow path, regardless of the selected userland
mitigations.

Comparing the slow path measurements to the fast path measurements, we observe
that the fast path measurements exhibit a reduced latency in comparison to the
slow path. Also, variance is significantly lower for the fast path
measurements. However, these effects were to be expected. Again, we do not see
significant impact of userland security features. Suprisingly, \grsec{} does
not have a significant impact on fast path latency either. Only in conjunction
with additional userland mitigations we see an increase in measurement result
variance. This suggests an interdependence of performance bottlenecks in userland
code that only surface if the binary is run on a \grsec{} enabled kernel.

\noindent \textbf{Summary:} Our measurements demonstrate that especially
userland mitigations do not have a significant impact on OvS forwarding
performance.  The use of \grsec{} kernel patches however does entail a
notable performance overhead. However, due to OvS being regularly present on
computation nodes and hypervisor systems, the overall system impact of \grsec{}
makes its use outside of dedicated network systems highly unlikely. On the 
other hand this also means that the kernel and fast path components of OvS
can certainly not be assumed to be protected by the various measures offered 
by \grsec{}.

\section{Discussion}
\label{sec:discussion}

While analyzing existing paradigms for SDN-, virtual switch-, data plane- and
control plane security, we have identified a new attack vector: on the
virtualized data plane.
Indeed, so far, the data plane has been of limited
interest with research focussing
on the control plane~\cite{hong2015poisoning,matsumoto2014fleet,dhawan2015sphinx} or forwarding issues in the data plane\cite{kamisinski2015flowmon,lee2006secure}.
We however, demonstrate
that the way virtual switches are run---namely privileged and
physically co-located with other critical components---they are suspectible to
attackers with limited time and money as well. This specifically regards cloud
tenants.  Given these insights, we were able to draft a new, weaker threat model
for infrastructures incorporating virtual switches.  Following this, we were
able to quickly identify an issue that allowed us to fully compromise a cloud
setup from within a single tenant machine.  To identify this vulnerability,
limited time and effort was required: we simply used standard vulnerability
scanning techniques, i.e., fuzzing.  Thus, the identified attack is not only
\emph{cheap} to execute by a \emph{non-sophisticated attacker}, but was also
easy to find. We only had to fuzz less than \emph{2$\%$} of the ovs-vswitchd
execution paths in order to find an exploitable vulnerability for
\emph{remote code execution}.

Hence, with this paper, we question existing assumptions and threat
models for (virtual) SDN switches as well as for the trade-off between data
plane and control plane.  Our threat model is worrisome, given that virtual switches such
as OvS are quite popular among cloud operating systems.
IT, telecommunications, academia and research, and finance organizations
are the majority
adopters of cloud operating systems such as OpenStack~\cite{ossurvey}.

The identified vulnerabilities can be leveraged to harm
essential services of the cloud operating system \os, including managed compute
resources (Hypervisors and Guest VMs), image management (the images VMs use to
boot-up), block storage (data storage), network management (virtual networks
between Hypervisors and Guest VMs), for the dashboard and web UI (in order to
manage the various resources from a centralized location), identity managment
(of the adminstrators and tenants), etc.

However, we have also observed that existing software-based countermeasures such as
stack canaries and PIE
effectively reduce the attack surface on the virtual switch.
They deter the attacker from exploiting
stack buffer overflows for remote code execution.
While in case
of kernel based countermeasures (using grsecurity),
this may come at a performance cost (especially in terms of latency),
our measurement results demonstrate that the performance overheads
of user-space countermeasures are negligible.

\subsection{Security Assumptions vs Performance}

As outlined above, our contributions should have a direct impact on how we approach
security for virtual switches and SDN platforms.  So far, a recurring theme for
virtual switches has been design and
performance~\cite{180182,Rizzo:2012:VSE:2413176.2413185,pfaff2015design}.  We
argue that the missing privilege separation and trust placed in virtual
switches are key-issues that needs to be addressed in order to achieve security
in SDN.

So far, one promising approach exists that
eliminates the hypervisor attack surface in the context of a cloud environment
by Szefer et al.~\cite{Szefer:2011:EHA:2046707.2046754}.
The hypervisor disengages itself from the guest VM, thereby, giving
the VM direct access to the hardware (e.g., NIC).
While such an approach protects the host running the virtual switch from full
compromise, the issue we raised on south-bound security remains open.

Due to the bi-directional nature of communication in SDN and virtual switching
environments, an attacker obtains direct access to control plane communication,
as soon as a switch is compromised. The consequences of this may be more dire
and complex in the context of 5G networks, where SDN is envisioned to control
programmable base stations and packet gateways~\cite{6812298}.  The base
stations are attacker facing entities in cellular networks. Therefore,
appropriate measures must be taken to ensure that compromising the base station
(data plane) does not lead to the compromise of the cellular network's control
plane.

In terms of software security, we find that existing security features, like
stack canaries,
may not be present for critical functions due to compiler optimizations.
Countermeasures such as PIE are not compiled for all packages
shipped by the operating system or vendor. This is important
because a major fraction of cloud operating systems' users simply
use the default packages~\cite{ossurvey}.

Our preliminary
performance measurements indicate that the overhead of unconditional stack
canaries and PIE together is acceptable for OvS. Hence, given the ease with
which we found opportunities for exploiting a virtual switch, adopting those
measures should be urgently done by maintainers and developers alike.

Note that the user-space mitigations would already have been sufficient to
mitigate the issues we found. In fact, the OvS user-space has, due to the
prelevance of packet \emph{parsing}~\cite{singaravelu2006reducing} there, a
much larger attack surface. 
While the evaluated userland mitigations did not introduce significant
overhead, applying \grsec{} in our evaluation led to significant impact. This,
again, highlights that clean coding and slim and well audited design are
crucial for the kernel-space parts (fast path) of virtual switches, as existing
mitigation techniques can not be easily applied there. While, e.g., clang
supports safe stack, it is not the officially supported compiler for the Linux
kernel. Hence, large distributions do not compile the Linux kernel with clang.

\subsection{Packet Processors Facing the Attacker}
A key feature of packet processing and forwarding systems in an SDN context 
is their ability to make forwarding decisions based on stateful information 
about a packet collected from all the network layers. This naturally also means
that such a system---of which virtual switches are an instance---has to
parse the protocols in unintended ways or on protocols from layers usually 
far beyond its actual layer of operation blurring the functionality between
switching and routing.

In this paper, the root-cause of one of the issues stems from handling the MPLS label in an
unintended manner. This parsing is done to derive additional information to
perform more efficient packet forwarding. MPLS, as described in
RFC~3031~\cite{rfc3031} and RFC~3032~\cite{rfc3032}
does not specify how to parse the whole label stack. Instead,
it specifies that when a labelled packet arrives, only the
top label must be used to make a forwarding decision.
However, in the case of OvS, all the labels in the
packet were parsed (beyond the supported limit) leading to a buffer overflow.
Security techniques such as explicitly indicating the size of the
label stack in the Shim header may not be acceptable as
from a networking perspective that is not required.

Similarly, it makes sense to parse higher layer protocol information in data
plane systems to request more efficient forwarding decisions from a controller.
Yet, the same problem arises if a data plane system attempts to parse higher
layer information in a single stream of packets. As soon as a data plane
system implements a parser for a protocol it is immediately susceptible to the same attack
surface as any daemon for that protocol.
Instead, the attack surface for the data plane
system rises indefinitely with each new protocol being parsed.

A possible method to mitigate these conceptual issues can be found in
a secure
packet processor architecture as suggested by
Chasaki~et~al.~\cite{chasaki2010design}:
monitor the control-flow
of the packet processor in hardware
and if any deviation from the known norm
occurs, to restore the processor to the
correct state.
However, the specific
approach outlined by Chasaki~et~al. is limited by the requirement to
be
implemented in hardware.
Furthermore, with protocol independent programmable packet processors
gaining momentum~\cite{barefoot,bosshart2014p4},
our findings highlight the consequences of
vulnerable packet processors.

\section{Related Work}
\label{sec:literature}
Attacks in the cloud have been demonstrated by a few researchers.
Ristenpart et al.~\cite{ristenpart2009hey} demonstrated how
an attacker can co-locate its VM with a target VM and then
steal the target's information.
We note this work is orthogonal to ours in that their objective was
co-locating their VM with the target VM and then stealing
that VMs information, while our work focusses on compromising
the server itself and extending that to all the other servers
in the cloud.
Costin et al.~\cite{costin2015all} examined the security of the
web-based interfaces offered by cloud providers.
Multiple vulnerabilities were exposed as a contribution
as well as possible attacks.
Wu et al.~\cite{wu2010network} assess
the network security of VMs in cloud computing.
The authors address the sniffing
and spoofing attacks a VM can carry out in a virtual
network and recommend placing a
firewall in the virtual network that prevents
such attacks.

Ristov et al.~\cite{ristov2013openstack} investigated 
the security of a default $\os$ deployment.
They show that it is vulnerable from
the inside rather than the outside. 
In the $\os$ security guide~\cite{openstackSecurityGuide},
it is mentioned that $\os$ is
inherently vulnerable due to the
bridged domains (Public and Management APIs,
Data and Management for a server, etc.).
Grobauer et al.~\cite{grobauer2011understanding}
take a general approach in classifying the possible
vulnerabilities
in 
cloud computing, and in doing so,
address the fact that the communication network
is vulnerable. 
However, there is no mention that the infrastructure that
enables the virtual networks can be vulnerable.
Porez-Botero et al.~\cite{perez2013characterizing}
characterize the possible
hypervisor vulnerabilities and state Network/IO as one.
However, they did
not find any known network based vulnerabilities at the time.

At the heart of the software-defined networking paradigm,
lies its support for formal policy specification and verification: it is generally believed that SDN has the potential to render computer networking more verifiable~\cite{header-space,veriflow} and even secure~\cite{fortnox,avantguard}.
However, researchers have recently also started to discover security threats in SDN. For example, Kloti et al.~\cite{kloti-stride} report on a STRIDE threat analysis of OpenFlow, and Kreutz et al.~\cite{Kreutz}
survey several threat vectors that may enable the exploitation of SDN vulnerabilities.

While much research went into designing more robust and secure SDN control planes~\cite{stn,nice,porras2015securing},
less published work exists on the issue of malicious switches (and data planes)~\cite{sonchack2016enabling,avantguard}.
However, the threat model introduced by an \emph{unreliable south-bound interface}, in which 
switches or routers do not behave as expected, but rather are malicious, is not new~\cite{spook,chi2015detect,dhawan2015sphinx,hong2015poisoning}.
%
%
In particular, national security agencies are reported to
have bugged networking equipment~\cite{snowdencisco}
and networking vendors have left
backdoors open~\cite{huawei,cryptoeprint:2016:376,netisbackdoor}.
However, in this paper we demonstrate that a weak
(resource constrained and unsophisticated)
attacker can impose serious damage:
compromise services far beyond the buggy virtual switch, and beyond simple 
denial-of-service attacks
(but affecting also, e.g., confidentiality and logical isolations). 


A closely related work on software switches is
by Chasaki et al.~\cite{chasaki2012attacks, chasaki2010design}
who uncover buffer overflow vulnerabilities 
and propose a secure packet processor 
to preserve the control flow of the packet processor
of the \emph{Click} software switch.
Additionally, Dobrescu et al.~\cite{dobrescu2014software} developed
a data plane verification tool to prove a
crash-free property of the \emph{Click} software switch's
data plane.

To the best of our knowledge, however, our work is the first to 
point out, characterize and demonstrate,
in a systematic manner, the severe vulnerabilities 
introduced in virtual switches used in cloud SDN deployments.

\section{Conclusion}
\label{sec:conclusion}
In this paper, we presented an 
analysis on how virtualization and
centralization of modern computer networks introduce new attack surfaces and
exploitation opportunities in and from the data plane. We 
demonstrated how even a simple, low-resource attacker can inflict serious harm
to distributed network systems.

Our key contribution is the realization that software defined networks in general and
virtual switches in particular suffer from conceptional challenges that have
not yet been sufficiently addressed:
\begin{enumerate}
	\item Insufficient privilege seperation in virtual switches.
	\item A virtualized and hardware co-located dataplane.
	\item Logical centralization and bi-directional communication in SDN.
	\item Support for extended protocol parsers.
\end{enumerate}


Following our analysis we derived a simplified attacker model for data plane
systems, which should be adapted. Furthermore, we
applied this threat model by performing attacks following its assumptions
and capabilities.
Subsequently, we were able to easily find and exploit a vulnerability in a virtual
switch, OvS, applying well known fuzzing techniques to its code-base.
With the exploit, we were able to fully take over a cloud setup (OpenStack)
within
a couple of minutes.

Our empirical experiments on the performance impact of
various software countermeasures on a virtual switch
debunks the myth that
hardened software is necessarily slow.
Instead they should be frequently adopted,
as they effectively reduce the attack surface on the virtual switch
while
their performance overhead in user-space is
negligible.
As our computer networks evolve and
networking concepts are shared across domains,
e.g., SDN being envisioned in 5G networks,
extensive work should be directed towards
privilege separation for virtual switches,
securing the data plane from attacks
and
designing secure packet processors.

\section*{Acknowledgements} The authors would like to express their gratitude
towards the German \emph{Bundesamt f{\"u}r Sicherheit in der Informationstechnik},
for sparking the authors' interest in SDN security. This work was partially
supported by the Danish Villum Foundation project ``ReNet'',  by BMBF
(Bundesministerium f{\"u}r Bildung und Forschung) Grant KIS1DSD032 (Project
Enzevalos) and by Leibniz Price project funds of DFG/German Research Foundation
(FKZ FE 570/4-1).  Furthermore, we would like to thank Jan Nordholz, Julian
Vetter, and Robert Buhren for their helpful discussions on the software
countermeasures. We would also like to thank the security team at Open vSwitch
for acknowledging our work in a timely and responsible manner.




\bibliographystyle{IEEEtranS}

\balance
\bibliography{master}

\begin{thebibliography}{10}
\providecommand{\url}[1]{#1}
\csname url@samestyle\endcsname
\providecommand{\newblock}{\relax}
\providecommand{\bibinfo}[2]{#2}
\providecommand{\BIBentrySTDinterwordspacing}{\spaceskip=0pt\relax}
\providecommand{\BIBentryALTinterwordstretchfactor}{4}
\providecommand{\BIBentryALTinterwordspacing}{\spaceskip=\fontdimen2\font plus
\BIBentryALTinterwordstretchfactor\fontdimen3\font minus
  \fontdimen4\font\relax}
\providecommand{\BIBforeignlanguage}[2]{{%
\expandafter\ifx\csname l@#1\endcsname\relax
\typeout{** WARNING: IEEEtranS.bst: No hyphenation pattern has been}%
\typeout{** loaded for the language `#1'. Using the pattern for}%
\typeout{** the default language instead.}%
\else
\language=\csname l@#1\endcsname
\fi
#2}}
\providecommand{\BIBdecl}{\relax}
\BIBdecl

\bibitem{osnetworkguide}
``Openstack networking-guide deployment scenarios,''
  \url{http://docs.openstack.org/liberty/networking-guide/deploy.html},
  accessed: 02-06-2016.

\bibitem{gadgettool}
``Ropgadget tool,''
  \url{https://github.com/JonathanSalwan/ROPgadget/tree/master}, accessed:
  02-06-2016.

\bibitem{ciscoVNlink}
``Cisco {VN-Link}: Virtualization-aware networking,'' White paper, 2009.

\bibitem{huawei}
``Huawei hg8245 backdoor and remote access,''
  \url{http://websec.ca/advisories/view/Huawei-web-backdoor-and-remote-access},
  2013, accessed: 27-01-2017.

\bibitem{netisbackdoor}
``Netis routers leave wide open backdoor,''
  \url{http://blog.trendmicro.com/trendlabs-security-intelligence/netis-routers-leave-wide-open-backdoor/},
  2014, accessed: 27-01-2017.

\bibitem{snowdencisco}
``Snowden: The {NSA} planted backdoors in cisco products,''
  \url{http://www.infoworld.com/article/2608141/internet-privacy/snowden--the-nsa-planted\\-backdoors-in-cisco-products.html},
  2014, accessed: 27-01-2017.

\bibitem{barefoot}
``{Barefoot Networks},'' \url{https://www.barefootnetwork.com/}, 2016.

\bibitem{openstackSecurityGuide}
``{OpenStack Security Guide},'' \url{http://docs.openstack.org/security-guide},
  2016, accessed: 27-01-2017.

\bibitem{Abadi05}
M.~Abadi, M.~Budiu, U.~Erlingsson, and J.~Ligatti, ``Control-flow integrity,''
  in \emph{Proc. ACM Conference on Computer and Communications Security (CCS)},
  2005, pp. 340--353.

\bibitem{commodity}
M.~Al-Fares, A.~Loukissas, and A.~Vahdat, ``A scalable, commodity data center
  network architecture,'' in \emph{Proc. ACM SIGCOMM}, 2008, pp. 63--74.

\bibitem{spook}
M.~Antikainen, T.~Aura, and M.~S{\"a}rel{\"a}, ``Spook in your network:
  Attacking an sdn with a compromised openflow switch,'' in \emph{Secure IT
  Systems: 19th Nordic Conference, NordSec 2014, Troms{\o}, Norway, October
  15-17, 2014, Proceedings}.\hskip 1em plus 0.5em minus 0.4em\relax Springer
  International Publishing, 2014, pp. 229--244.

\bibitem{rfc1812}
\BIBentryALTinterwordspacing
F.~Baker, ``{Requirements for IP Version 4 Routers},'' RFC 1812 (Proposed
  Standard), Internet Engineering Task Force, June 1995, updated by RFCs 2644,
  6633. [Online]. Available: \url{http://www.ietf.org/rfc/rfc1812.txt}
\BIBentrySTDinterwordspacing

\bibitem{blake2009survey}
G.~Blake, R.~G. Dreslinski, and T.~Mudge, ``A survey of multicore processors,''
  \emph{IEEE Signal Processing Magazine}, vol.~26, no.~6, pp. 26--37, November
  2009.

\bibitem{bosshart2014p4}
P.~Bosshart, D.~Daly, and G.~e.~a. Gibb, ``P4: Programming protocol-independent
  packet processors,'' \emph{ACM Computer Communication Review (CCR)}, vol.~44,
  no.~3, pp. 87--95, July 2014.

\bibitem{stn}
M.~Canini, P.~Kuznetsov, D.~Levin, and S.~Schmid, ``A distributed and robust
  sdn control plane for transactional network updates,'' in \emph{Proc. IEEE
  INFOCOM}, April 2015, pp. 190--198.

\bibitem{nice}
M.~Canini, D.~Venzano, P.~Pere{\v s}{\'\i}ni, D.~Kosti{\'c}, and J.~Rexford,
  ``A nice way to test openflow applications,'' in \emph{Proc. Usenix Symposium
  on Networked Systems Design and Implementation (NSDI)}, 2012, pp. 127--140.

\bibitem{Casado:2010:VNF:1921151.1921162}
M.~Casado, T.~Koponen, R.~Ramanathan, and S.~Shenker, ``Virtualizing the
  network forwarding plane,'' in \emph{Proc. ACM CoNEXT Workshop on
  Programmable Routers for Extensible Services of Tomorrow}, 2010, pp.
  8:1--8:6.

\bibitem{chasaki2010design}
D.~Chasaki and T.~Wolf, ``Design of a secure packet processor,'' in \emph{Proc.
  ACM/IEEE Architectures for Networking and Communication Systems (ANCS)}, Oct
  2010, pp. 1--10.

\bibitem{cryptoeprint:2016:376}
S.~Checkoway \emph{et~al.}, ``A systematic analysis of the juniper dual ec
  incident,'' Cryptology ePrint Archive, Report 2016/376, 2016.

\bibitem{chen2005}
S.~Chen, J.~Xu, E.~C. Sezer, P.~Gauriar, and R.~K. Iyer, ``Non-control-data
  attacks are realistic threats.'' in \emph{Proc. Usenix Security Symp.},
  vol.~5, 2005.

\bibitem{chi2015detect}
P.-W. Chi, C.-T. Kuo, J.-W. Guo, and C.-L. Lei, ``How to detect a compromised
  sdn switch,'' in \emph{Network Softwarization (NetSoft), 2015 1st IEEE
  Conference on}, April 2015, pp. 1--6.

\bibitem{6812298}
W.~H. Chin, Z.~Fan, and R.~Haines, ``Emerging technologies and research
  challenges for 5g wireless networks,'' \emph{IEEE Wireless Communications},
  vol.~21, no.~2, pp. 106--112, April 2014.

\bibitem{costin2015all}
A.~Costin, ``All your cluster-grids are belong to us: Monitoring the
  (in)security of infrastructure monitoring systems,'' in \emph{Proc. IEEE
  Communications and Network Security (CNS)}, Sept 2015, pp. 550--558.

\bibitem{Cowan98}
C.~Cowan \emph{et~al.}, ``Stackguard: Automatic adaptive detection and
  prevention of buffer-overflow attacks,'' in \emph{Proc. Usenix Security
  Symp.}, 1998, pp. 5--5.

\bibitem{chasaki2012attacks}
{D. Chasaki} and T.~Wolf, ``Attacks and defenses in the data plane of
  networks,'' \emph{Proc. IEEE/IFIP Transactions on Dependable and Secure
  Computing (DSN)}, vol.~9, no.~6, pp. 798--810, Nov 2012.

\bibitem{dhawan2015sphinx}
M.~Dhawan, R.~Poddar, K.~Mahajan, and V.~Mann, ``Sphinx: Detecting security
  attacks in software-defined networks.'' in \emph{Proc. Internet Society
  Symposium on Network and Distributed System Security (NDSS)}, 2015.

\bibitem{dobrescu2014software}
M.~Dobrescu and K.~Argyraki, ``Software dataplane verification,'' in
  \emph{Proc. Usenix Symposium on Networked Systems Design and Implementation
  (NSDI)}, April 2014, pp. 101--114.

\bibitem{opennf}
A.~Gember-Jacobson \emph{et~al.}, ``Opennf: Enabling innovation in network
  function control,'' in \emph{Proc. ACM SIGCOMM}, 2014, pp. 163--174.

\bibitem{greenberg2015sdn}
A.~Greenberg, ``Sdn for the cloud,'' in \emph{Keynote in the 2015 ACM SIGCOMM},
  2015.

\bibitem{grobauer2011understanding}
B.~Grobauer, T.~Walloschek, and E.~Stocker, ``Understanding cloud computing
  vulnerabilities,'' \emph{Proc. IEEE Security \& Privacy (S\&P)}, vol.~9,
  no.~2, pp. 50--57, March 2011.

\bibitem{hong2015poisoning}
S.~Hong, L.~Xu, H.~Wang, and G.~Gu, ``Poisoning network visibility in
  software-defined networks: New attacks and countermeasures.'' in \emph{Proc.
  Internet Society Symposium on Network and Distributed System Security
  (NDSS)}, 2015.

\bibitem{jacobson1988congestion}
V.~Jacobson, ``Congestion avoidance and control,'' in \emph{ACM Computer
  Communication Review (CCR)}, vol.~18, no.~4, 1988, pp. 314--329.

\bibitem{kamath2010edge}
D.~Kamath \emph{et~al.}, ``Edge virtual bridge proposal, version 0. rev. 0.1,''
  \emph{Apr}, vol.~23, pp. 1--72, 2010.

\bibitem{kamisinski2015flowmon}
A.~Kamisi{\'n}ski and C.~Fung, ``Flowmon: Detecting malicious switches in
  software-defined networks,'' in \emph{Proc. ACM Workshop on Automated
  Decision making for Active Cyber Defense}, 2015, pp. 39--45.

\bibitem{header-space}
P.~Kazemian, G.~Varghese, and N.~McKeown, ``Header space analysis: Static
  checking for networks,'' in \emph{Proc. Usenix Symposium on Networked Systems
  Design and Implementation (NSDI)}, 2012, pp. 113--126.

\bibitem{veriflow}
A.~Khurshid \emph{et~al.}, ``Veriflow: Verifying network-wide invariants in
  real time,'' in \emph{Proc. Usenix Symposium on Networked Systems Design and
  Implementation (NSDI)}, 2013, pp. 15--27.

\bibitem{kloti-stride}
R.~Kl{\"o}ti, V.~Kotronis, and P.~Smith, ``Openflow: A security analysis,'' in
  \emph{Proc. IEEE International Conference on Network Protocols (ICNP)}, Oct
  2013, pp. 1--6.

\bibitem{Kreutz}
D.~Kreutz, F.~M. Ramos, and P.~Verissimo, ``Towards secure and dependable
  software-defined networks,'' in \emph{Proc. ACM Workshop on Hot Topics in
  Software Defined Networking (HotSDN)}, 2013, pp. 55--60.

\bibitem{186159}
V.~Kuznetsov \emph{et~al.}, ``Code-pointer integrity,'' in \emph{Proc. Usenix
  Symposium on Operating Systems Design and Implementation (OSDI)}, October
  2014, pp. 147--163.

\bibitem{lee2006secure}
S.~Lee, T.~Wong, and H.~S. Kim, ``Secure split assignment trajectory sampling:
  A malicious router detection system,'' in \emph{Proc. IEEE/IFIP Transactions
  on Dependable and Secure Computing (DSN)}, 2006, pp. 333--342.

\bibitem{reading}
{Light Reading}, ``Alcatel-lucent joins virtual router race,''
  {http://www.lightreading.com/nfv/nfv-elements/alcatel-lucent-joins-virtual-router-race/d/d-id/712004},
  2014.

\bibitem{martins2014clickos}
J.~Martins \emph{et~al.}, ``Clickos and the art of network function
  virtualization,'' in \emph{Proc. Usenix Symposium on Networked Systems Design
  and Implementation (NSDI)}, 2014, pp. 459--473.

\bibitem{matsumoto2014fleet}
S.~Matsumoto, S.~Hitz, and A.~Perrig, ``Fleet: Defending sdns from malicious
  administrators,'' in \emph{Proc. ACM Workshop on Hot Topics in Software
  Defined Networking (HotSDN)}, 2014, pp. 103--108.

\bibitem{PaX01}
{PaX}, ``{The Guaranteed End of Arbitrary Code Execution},''
  \url{https://grsecurity.net/PaX-presentation.ppt}.

\bibitem{payer2012}
M.~Payer, ``{Too much PIE is bad for performance},''
  \url{http://e-collection.library.ethz.ch/eserv/eth:5699/eth-5699-01.pdf},
  2012, accessed: 27-01-2017.

\bibitem{perez2013characterizing}
D.~Perez-Botero, J.~Szefer, and R.~B. Lee, ``Characterizing hypervisor
  vulnerabilities in cloud computing servers,'' in \emph{Proc. ACM Workshop on
  Security in Cloud Computing}, 2013, pp. 3--10.

\bibitem{pettit2010virtual}
J.~Pettit, J.~Gross, B.~Pfaff, M.~Casado, and S.~Crosby, ``Virtual switching in
  an era of advanced edges,'' Technical Report.

\bibitem{pfaff2009extending}
B.~Pfaff, J.~Pettit, K.~Amidon, M.~Casado, T.~Koponen, and S.~Shenker,
  ``Extending networking into the virtualization layer.'' in \emph{Proc. ACM
  Workshop on Hot Topics in Networks (HotNETs)}, 2009.

\bibitem{pfaff2015design}
B.~Pfaff, J.~Pettit, T.~Koponen \emph{et~al.}, ``The design and implementation
  of {Open} {vSwitch},'' in \emph{Proc. Usenix Symposium on Networked Systems
  Design and Implementation (NSDI)}, May 2015, pp. 117--130.

\bibitem{fortnox}
P.~Porras, S.~Shin, V.~Yegneswaran, M.~Fong, M.~Tyson, and G.~Gu, ``A security
  enforcement kernel for {OpenFlow} networks,'' in \emph{Proc. ACM Workshop on
  Hot Topics in Software Defined Networking (HotSDN)}, 2012, pp. 121--126.

\bibitem{porras2015securing}
P.~Porras, S.~Cheung, M.~Fong, K.~Skinner, and V.~Yegneswaran, ``Securing the
  software-defined network control layer,'' in \emph{Proc. Internet Society
  Symposium on Network and Distributed System Security (NDSS)}, 2015.

\bibitem{180182}
K.~K. Ram \emph{et~al.}, ``Hyper-switch: A scalable software virtual switching
  architecture,'' in \emph{Usenix Annual Technical Conference (ATC)}, 2013, pp.
  13--24.

\bibitem{ristenpart2009hey}
T.~Ristenpart, E.~Tromer, H.~Shacham, and S.~Savage, ``Hey, you, get off of my
  cloud: Exploring information leakage in third-party compute clouds,'' in
  \emph{Proc. ACM Conference on Computer and Communications Security (CCS)},
  2009, pp. 199--212.

\bibitem{ristov2013openstack}
S.~Ristov, M.~Gusev, and A.~Donevski, ``Openstack cloud security
  vulnerabilities from inside and outside,'' Technical Report, pp. 101--107,
  2013.

\bibitem{Rizzo:2012:VSE:2413176.2413185}
L.~Rizzo and G.~Lettieri, ``{VALE}, a switched ethernet for virtual machines,''
  in \emph{Proc. ACM CoNEXT}, 2012, pp. 61--72.

\bibitem{roemer2012return}
R.~Roemer, E.~Buchanan, H.~Shacham, and S.~Savage, ``Return-oriented
  programming: Systems, languages, and applications,'' \emph{ACM Trans. on
  Information and System Security (TISSEC)}, vol.~15, no.~1, pp. 2:1--2:34,
  March 2012.

\bibitem{rfc3032}
\BIBentryALTinterwordspacing
E.~Rosen, D.~Tappan, G.~Fedorkow, Y.~Rekhter, D.~Farinacci, T.~Li, and
  A.~Conta, ``{MPLS Label Stack Encoding},'' RFC 3032 (Proposed Standard),
  Internet Engineering Task Force, January 2001, updated by RFCs 3443, 4182,
  5332, 3270, 5129, 5462, 5586, 7274. [Online]. Available:
  \url{http://www.ietf.org/rfc/rfc3032.txt}
\BIBentrySTDinterwordspacing

\bibitem{rfc3031}
\BIBentryALTinterwordspacing
E.~Rosen, A.~Viswanathan, and R.~Callon, ``{Multiprotocol Label Switching
  Architecture},'' RFC 3031 (Proposed Standard), Internet Engineering Task
  Force, January 2001, updated by RFCs 6178, 6790. [Online]. Available:
  \url{http://www.ietf.org/rfc/rfc3031.txt}
\BIBentrySTDinterwordspacing

\bibitem{schuchard2010losing}
M.~Schuchard \emph{et~al.}, ``Losing control of the internet: using the data
  plane to attack the control plane,'' in \emph{Proc. ACM Conference on
  Computer and Communications Security (CCS)}, 2010, pp. 726--728.

\bibitem{opensdwn}
J.~Schulz-Zander, C.~Mayer, B.~Ciobotaru, S.~Schmid, and A.~Feldmann,
  ``Opensdwn: Programmatic control over home and enterprise wifi,'' in
  \emph{Proc. ACM Symposium on Software Defined Networking Research (SOSR)},
  2015, pp. 16:1--16:12.

\bibitem{manifesto}
V.~Sekar \emph{et~al.}, ``The middlebox manifesto: Enabling innovation in
  middlebox deployment,'' in \emph{Proc. ACM Workshop on Hot Topics in Networks
  (HotNETs)}, 2011, pp. 21:1--21:6.

\bibitem{someone}
J.~Sherry \emph{et~al.}, ``Making middleboxes someone else's problem: Network
  processing as a cloud service,'' vol.~42, no.~4.\hskip 1em plus 0.5em minus
  0.4em\relax ACM, August 2012, pp. 13--24.

\bibitem{fresco}
S.~Shin, P.~Porras, V.~Yegneswaran, M.~Fong, G.~Gu, and M.~Tyson, ``Fresco:
  Modular composable security services for software-defined networks,'' in
  \emph{Proc. Internet Society Symposium on Network and Distributed System
  Security (NDSS)}, 2013.

\bibitem{avantguard}
S.~Shin, V.~Yegneswaran, P.~Porras, and G.~Gu, ``{AVANT-GUARD}: Scalable and
  vigilant switch flow management in software-defined networks,'' in
  \emph{Proc. ACM Conference on Computer and Communications Security (CCS)},
  2013, pp. 413--424.

\bibitem{singaravelu2006reducing}
L.~Singaravelu, C.~Pu, H.~H{\"a}rtig, and C.~Helmuth, ``Reducing tcb complexity
  for security-sensitive applications: Three case studies,'' in \emph{ACM
  SIGOPS Operating System Review}, vol.~40, no.~4.\hskip 1em plus 0.5em minus
  0.4em\relax ACM, 2006, pp. 161--174.

\bibitem{sonchack2016enabling}
J.~Sonchack, A.~J. Aviv, E.~Keller, and J.~M. Smith, ``Enabling practical
  software-defined networking security applications with {OFX},'' in
  \emph{Proc. Internet Society Symposium on Network and Distributed System
  Security (NDSS)}, 2016.

\bibitem{Szefer:2011:EHA:2046707.2046754}
J.~Szefer \emph{et~al.}, ``Eliminating the hypervisor attack surface for a more
  secure cloud,'' in \emph{Proc. ACM Conference on Computer and Communications
  Security (CCS)}, 2011, pp. 401--412.

\bibitem{179731}
{T. Koponen et al.}, ``Network virtualization in multi-tenant datacenters,'' in
  \emph{11th USENIX Symposium on Networked Systems Design and Implementation},
  2014.

\bibitem{ossurvey}
H.~J. Tretheway \emph{et~al.}, ``A snapshot of openstack users' attitudes and
  deployments.'' \emph{Openstack User Survey}, Apr 2016.

\bibitem{wu2010network}
H.~Wu \emph{et~al.}, ``Network security for virtual machine in cloud
  computing,'' in \emph{Proc. IEEE Conference on Computer Sciences and
  Convergence Information Technology}, Nov 2010, pp. 18--21.

\end{thebibliography}
\balance

\end{document}